\documentclass[12pt]{article}
\usepackage[dvips]{graphicx}
\usepackage{epsfig,cite}
\usepackage {amsmath}
\usepackage{amssymb}
\usepackage{slashed}
\usepackage{simplewick}

\def\mpl{M_{P}}

\def\beq{\begin{equation}}
\def\eeq{\end{equation}}
\def\beqn{\begin{eqnarray}}
\def\eeqn{\end{eqnarray}}

\begin{document}

\begin{titlepage}
\pagestyle{empty}
\baselineskip=21pt
\rightline{UMN--TH-2906/10}
\rightline{FTPI--MINN--10/14}
\vskip +0.4in
\begin{center}
{\large {\bf The gravitino coupling to broken gauge theories applied to the MSSM}}

\end{center}
\begin{center}
\vskip +0.2in
{\bf Feng Luo}$^{1}$, {\bf Keith~A.~Olive}$^{1,2}$,
and {\bf Marco Peloso}$^{1}$
\vskip 0.3in
{\small {\it
$^1${School of Physics and Astronomy, \\
University of Minnesota, Minneapolis, MN 55455, USA}\\
$^2${William I. Fine Theoretical Physics Institute, \\
University of Minnesota, Minneapolis, MN 55455, USA}\\
}}

\vskip 0.3in

{\bf Abstract}
\end{center}

{\small
We consider gravitino couplings in theories with broken gauge symmetries.
In particular, we compute the single gravitino production cross section in
$W^+ W^-$ fusion processes.  Despite recent claims to the contrary,
we show that this process is always subdominant to gluon fusion
processes in the high energy limit.  The full calculation is performed
numerically; however, we give analytic expressions for the cross section in the supersymmetric
and electroweak limits.  We also confirm these results with the use of the effective theory
of goldstino interactions.}

\end{titlepage}

\section{Introduction}

One of the reasons that supersymmetric theories are the prime focus
for physics beyond the standard model, are their inherent ability to
be tested. In addition to its many more theoretical benefits such as
stabilization of the electroweak symmetry breaking scale
\cite{hierarchy} and unification of gauge couplings at high energy
\cite{GUT}, low energy supersymmetry \cite{MSSM} often predicts a
particle spectrum readily observable at colliders such as the LHC
\cite{baer}.  Indeed, in models where unification conditions are
placed at some high energy scale, such as in the constrained minimal
supersymmetric standard model (CMSSM), regions of parameter space
consistent with known phenomenological constraints at the 95 \% CL
are well within the expected reach of the LHC \cite{mc3}. If
$R$-parity is conserved, supersymmetry (SUSY) also predicts that the
lightest supersymmetric particle (LSP)
 is stable, thereby making it an excellent candidate for dark matter and if it is the neutralino \cite{EHNOS}, it is also potentially observable in direct detection experiments \cite{eosk5}.
 In this case, it is usually {\em assumed} that the gravitino is heavier than the neutralino,
and even then, additional assumptions must be made so that its decays in the early universe
do not upset the results of big bang nucleosynthesis \cite{decay,both}.

It is also quite possible that the gravitino is the LSP \cite{gdm1,EHNOS,gdm2}.
In the CMSSM, this will occur whenever the gravitino mass, $m_{3/2}$ is less than the lightest
standard model superpartner mass \cite{gdm2} making it subject to big bang nucleosynthesis constraints on the decays of the next to lightest supersymmetric particle (NLSP) \cite{gdmdecay,notchi}. Indeed, a gravitino LSP is quite common in models based on minimal supergravity \cite{vcmssm}.  Typically, one would expect gravitino masses of order the weak scale,
making direct detection of dark matter very unlikely.  There are nevertheless proposals
for detecting the long lived decays of a stau at the LHC \cite{bhry,noj}.

The possibility of a light gravitino precedes the MSSM \cite{Fayet,Fayet2} and
in models of gauge meditated supersymmetry breaking \cite{gm}, the gravitino may be significantly
lighter with masses as low as $10^{-5}$ -- $10^{-6}$ eV. While cosmological
constraints on these models may be derived \cite{gmcon}, there remains a broad mass
range for super-light gravitinos. In no-scale supergravity models \cite{noscale},
the gravitino mass is decoupled from the rest of the supersymmetric sparticle spectrum and may
be set to the Planck scale \cite{Ellis:1984bm}, or to the keV scale and below \cite{noscale2}.

The detection of very light gravitinos at colliders {\em is} in principle possible through the
decay of the NLSP \cite{lightdecay,dn} or through direct production at $e^+ e^-$ \cite{eeprod}
or hadron \cite{dn,hprod} colliders. This is possible, because, as we will see,
the gravitino couplings are inversely proportional to its mass, making very light
gravitinos  more readily accessible. The current {\em lower} bound on the
mass of a super-light gravitino
comes from LEP and is \cite{Achard:2003tx}
\beq
m_{3/2} > 1.35 \times 10^{-5} {\rm eV} ,
\label{lep}
\eeq
and this limit will surely be improved at the LHC.
The dominant processes affected by a light gravitino are expected to proceed through
the pair production of gluinos
\beq
p { \bar p} \to {\tilde g} {\tilde g}
\eeq
or through associated gravitino production with either squarks or gluinos
\beq
p {\bar p} \to {\tilde g} {\tilde G} \, , {\tilde q} {\tilde G}
\eeq

It is known that gravitino production processes suffer a breakdown of unitarity at
high energies due to the non-renormalizability of the super-gravity Lagrangian \cite{nonu}.
However, given the mass bound (\ref{lep}), unitarity is preserved through the TeV scale.
Recently,  it was claimed \cite{Ferrantelli:2007bx} that the breakdown of unitarity is
significantly more severe in theories with a broken gauge symmetry such as the Standard Model.
Indeed, it was claimed that in the high energy limit, the cross section for gravitino production
remains non-zero even in the limit of exact supersymmetry.
If true, this would imply that associated production of gravitinos through W boson fusion
would come to dominate at high energy when compared to gluon fusion (where the
gauge symmetry is unbroken).  Here, we will calculate the weak boson fusion
process leading to gravitino production and show that, contrary to the claims of
\cite{Ferrantelli:2007bx}, the cross section is well-behaved at high energy.

For very light gravitino masses, couplings of the gravitino to
matter are dominated by the goldstino and gluon fusion process will
therefore be proportional to $m_{\tilde g}^2/m_{3/2}^2$. Single
gravitino production through gluon fusion, $pp \rightarrow {\tilde
g} {\tilde G}$ was recently reconsidered in \cite{Klasen:2006kb},
where they found
\begin{equation}
\sigma_g \sim {\rm few} \times {\rm pb} \, \left( \frac{10^{-4} \, {\rm eV}}{m_{3/2}} \right)^2 \left( \frac{m_{susy}}{1~{\rm TeV}} \right)^2
\end{equation}
assuming mass spectra corresponding to SPS benchmark points 7 and 8 \cite{sps}
for which the gluino mass is 920 and 810 GeV respectively.

In contrast, neutralino pair production through $W^+ W^-$ fusion, $pp \rightarrow \chi^0 \chi^0$,
was considered in Ref. \cite{Cho:2006sx}.
Cross sections for the same SPS benchmark points $7$ and $8$ were found to be as high as $\sim 0.1 \, {\rm fb}$ (for the production of $\chi_2^0 \chi_2^0$). We would naively estimate that
single gravitino production would scale as
\begin{eqnarray}
&& \sigma_{pp \rightarrow  \chi^0 G}\equiv \sigma_W \sim  10^{-8} \left( \frac{m_{\rm susy} }{M_P m_{3/2} }\right)^2 \sim
10^{-4} \, {\rm pb} \,  \left( \frac{10^{-4} \, {\rm eV}}{m_{3/2}} \right)^2 \left( \frac{m_{susy}}{1~{\rm TeV}} \right)^2
\label{west}
\end{eqnarray}
where  $m_{\rm susy}$ is the typical mass of supersymmetric particles (close to the electroweak scale) and $\mpl$ is the reduced Planck mass $\mpl = 1/\sqrt{8\pi G_N}
 \approx 2.4 \times 10^{18}$ GeV , where $G_N$ is the Newton
gravitational constant.

\begin{figure}[h]
\centerline{
\includegraphics[width=0.6\textwidth,angle=-90]{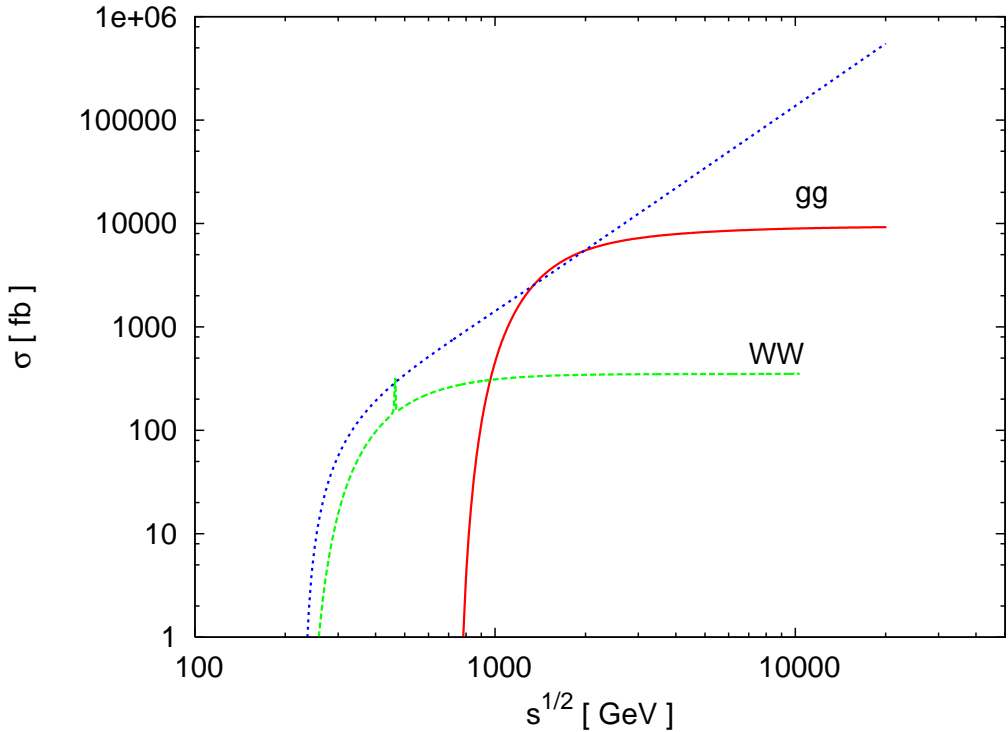}
}
\caption{Single gravitino production cross section as a function of center of mass energy.
The gluon fusion process leading to gravitino plus gluino is shown by the solid curve labelled gg.
The cross section for W fusion to gravitino plus neutralino predicted in \protect\cite{Ferrantelli:2007bx}
is shown by the dotted curve.  Our calculation for the same process is shown by the dashed curve
labelled WW. Choices for the supersymmetric parameters used are discussed in section \ref{sect:comp}. }
\label{fig:compare}
\end{figure}

The ratio between the gluon and $W$ fusion productions are therefore estimated to be
\begin{equation}
\frac{\sigma_W}{\sigma_g} \sim 10^{-4}
\end{equation}
If the claim in \cite{Ferrantelli:2007bx} was right, one would get an additional factor of  $s/m_{\rm susy}^2$:
\begin{equation}
\frac{\sigma_W}{\sigma_g} \vert_{\rm Ferrantelli} \sim 10^{-4} \left( \frac{s}{m_{\rm susy}^2} \right)
\label{ferrfact}
\end{equation}
where $s$ is the square of the center of mass energy in the collision between $W^+ W^-$. This ratio can in principle be of order one, or bigger.
In Fig.~\ref{fig:compare}, we show the qualitative behavior of the cross section for single gravitino
production in the symmetric case of gluon fusion (solid curve labelled gg), and in the broken case of
W boson fusion (dashed curve labelled WW). If eq. (\ref{ferrfact}) holds, W fusion process would come to dominate over gluonic ones (as shown by the dotted curve). Note that the cross sections shown
in the figure do not include necessary form factors and so do not represent $p$ $p$ cross sections.

Furthermore, we point out that the claim of
\cite{Ferrantelli:2007bx} is also in contradiction with the
equivalence theorem, which states that the gravitino can be
effectively replaced by the goldstino \cite{Fayet} at energies much
greater than its mass. In calculations based on the equivalence
theorem, one uses the on-shell conservation of the supercurrent, to
which the goldstino is coupled to. If the result of
\cite{Ferrantelli:2007bx} was correct, it would imply that a broken
gauge theory contains some loopholes that invalidate the theorem,
and that would not allow one to use the equations of motion in
determining the couplings of the goldstino. This prompted us to
revisit this issue, and to provide an explicit proof of the theorem.
We do so by generalizing the calculation of  \cite{Lee:1998aw},
where the equivalence is shown at the level of $S-$matrix elements.
The calculation of  \cite{Lee:1998aw} is performed for an unbroken
U(1) theory, and the effective Lagrangian included a term which was
previously unnoticed. The result of \cite{Lee:1998aw} is confirmed
by \cite{Rychkov:2007uq} and the effective Lagrangian is related to
the soft SUSY breaking terms in the MSSM through an explicit use of
the equations of motion.  Here, we extend these derivations to
include the case of broken gauge symmetries. We make this relation
explicit in equations (\ref{eq:equivalence}) and (\ref{eq:non-der})
below, and prove it in Appendix \ref{app:equivalence}. It is
manifest from the proof that the theorem applies irrespectively to
whether the gauge symmetry is or is not broken.

The plan of the paper is as follows. In the next section, we write
out the interaction Lagrangian for the gravitino coupled to the MSSM
with broken electroweak symmetry. In anticipation of taking the high
energy limit, where we can replace the gravitino couplings with the
goldstino, we write out the explicit couplings of the goldstino to
the terms originating from the soft supersymmetry breaking
Lagrangian. In section \ref{sect:comp}, we outline our calculation
of the $W^+ W^- \to {\tilde G} \chi^0$ cross section.  While the
analytic expression for the cross section is too long to write out,
we do give analytical results in a couple of interesting limits.
First, we show that in the supersymmetric limit ($m_{3/2} \ll m_{\rm
susy} \ll m_W$), $\overline{\vert {\cal M} \vert ^2} \propto m_{\rm
susy}^2 s/m_{3/2}^2 M_P^2$ at high energy which would lead to a
cross section for gravitino production of the form given in eq.
(\ref{west}). We also consider the limit $m_{3/2} \ll m_W \ll m_{\rm
susy} $ and write out the analytical cross section at high energy,
which takes a similar form. We also comment on the detectability  of
gravitinos through this process in comparison with that of gluon
fusion.  Concluding remarks are given in section \ref{sect:conc}. We
also show explicitly our derivation of effective gravitino
Lagrangian in the appendix.

\section{Interaction Lagrangian for the gravitino with broken electroweak symmetry}
\label{sec:interactions}

The interactions vertices between a single gravitino and the MSSM fields are obtained from the interaction Lagrangian
\begin{eqnarray}
\label{eq:grav int L}
\mathcal{L}_\text{int} &=&- \, \frac{i}{\sqrt{2} \, M_P} \, {\bar \psi}_\mu \, S_{\rm MSSM}^\mu + {\rm h. \, c.}  \nonumber\\
&=&  -\frac{i}{\sqrt{2}\mpl} \left[
\mathcal D_{\nu}^{(\alpha)} \phi^{*i} \overline{\psi}_\mu \gamma^{\nu}
\gamma^{\mu} \chi^{i}_{L}
-\mathcal D_{\nu}^{(\alpha)} \phi^{i} \overline{\chi}^{i}_{L} \gamma^\mu \gamma^\nu \psi_\mu
\right] \nonumber \\&&
- \frac{i}{8\mpl}\overline{\psi}_\mu [\gamma^{\rho}\,,\gamma^{\sigma}]
\gamma^{\mu} \lambda^{(\alpha)} F_{\rho\sigma}^{(\alpha)}   \;\;,
\end{eqnarray}
where, following the notation of \cite{Pradlerthesis}, $\psi_\mu$
denotes the gravitino field, $\phi$ and $\chi_L$ the scalar and
fermion components of the chiral MSSM superfields, $F_{\rho\sigma}$
is the field strength of a gauge boson field, and $\lambda$ is the
corresponding gaugino. The indices $i$ and ($\alpha$) label
the chiral and gauge multiplets, respectively (notice that we are
implicitly summing over all the MSSM chiral and gauge multiplets).
The covariant derivative of a scalar field is
\begin{equation}
\mathcal D_{\mu}^{(\alpha)} \phi^{i} \equiv \partial_\mu \phi^{i} + i \, g_{(\alpha)} \, A_\mu^{(\alpha) a} \, \left( T^{(\alpha) a} \, \phi \right)^{i}
\end{equation}

In the first line of (\ref{eq:grav int L}),  $S_{\rm MSSM}^\mu$
denotes the contribution from the MSSM fields to the supercurrent
and contains only terms from the supersymmetric Lagrangian.
Specifically, under a supersymmetry transformation, any MSSM field
(of any spin) $\Phi_i$ transforms as $\Phi_i \rightarrow \Phi_i +
\delta \Phi_i $, while the supersymmetric part of the MSSM
Lagrangian transforms as $\mathcal{L}_{\rm susy} \rightarrow
\mathcal{L}_{\rm susy} + \partial_\mu K^\mu$. Then, the supercurrent
is
\begin{eqnarray}
\label{eq:supercurrent}
S_{\rm MSSM}^\mu &\equiv& \frac{\partial \mathcal{L}_{\rm susy}}{\partial \left( \partial_\mu \, \Phi_i \right)} \, \delta \Phi_i - K^\mu
\end{eqnarray}
The explicit expression for the supercurrent can be found for example in~\cite{Rychkov:2007uq}.

We want to single out the gravitino interactions that arise due to the breaking of the electroweak symmetry. Following \cite{GH1986}, we denote the two Higgs doublets as
\begin{equation}
H_1 \equiv \left( \begin{array}{c} H_1^1 \\ H_1^2 \end{array} \right) \;\;,\;\;
H_2 \equiv \left( \begin{array}{c} H_2^1 \\ H_2^2 \end{array} \right) \;\;,\;\;
\end{equation}
and we denote their vacuum expectation values (vevs) as $\langle H_1^1 \rangle \equiv v_1 \,,\,
\langle H_2^2 \rangle \equiv v_2 \,,\, \langle H_1^2 \rangle = \langle H_2^1 \rangle = 0$. We denote the corresponding higgsino fields as
\begin{equation}
{\tilde H}_{1 \, L}^1 \equiv P_L \, {\tilde H}_1 \;\;,\;\;
{\tilde H}_{1 \, L}^2 \equiv P_L \, {\tilde H}^- \;\;,\;\;
{\tilde H}_{2 \, L}^1 \equiv P_L \, {\tilde H}^+ \;\;,\;\;
{\tilde H}_{2 \, L}^2 \equiv P_L \, {\tilde H}_2 \;\;,\;\;
\end{equation}
where ${\tilde H}^- = \left( {\tilde H}^+ \right)^c$.

Then the interaction term we are interested in is
\begin{eqnarray}
{\cal L}_{\rm int } &\supset& - \frac{g}{M_P} \, {\bar \psi}_\mu \left[ v_1 \, W^{\mu +}  \, P_L {\tilde H}^- + v_2 \, W^{\mu -} \, \, P_L {\tilde H}^+
\right. \nonumber\\
&&\left. \quad\quad\quad\quad\quad\quad
+ \frac{Z^\mu}{\sqrt{2} \, \cos \theta_w} \,  \left( v_1 \, P_L {\tilde H}_1 - v_2 \, P_L {\tilde H}_2 \right) \right] + {\rm h. c.}
\label{int-propto-v}
\end{eqnarray}
(where we have used ${\bar \psi}_\mu \gamma^\mu = 0$, and where the
gauge fields are defined in the standard way, see for instance
\cite{MSSM}). We rewrite these interactions in terms of the chargino
(${\tilde \chi}_j \;, j = 1,\, 2$) and neutralino (${\tilde
\chi}^0_i \;, i =  1  ,\dots, 4$) mass eigenstates, using the
rotation formulae \cite{GH1986}
\begin{eqnarray}
&&P_L \, {\tilde H}^+ = V_{j2}^* \, P_L \, {\tilde \chi}_j  \;\;\;,\;\;\;
P_L \, {\tilde H}^- = U_{j2}^* \, P_L \, {\tilde \chi}_j^c  \;\;\;,\;\;\; \nonumber\\
&&P_L \, {\tilde H}_1 = N_{i\,3}^* \, P_L \, {\tilde \chi}^0_i
\;\;\;,\;\;\; P_L \, {\tilde H}_2 = N_{i\,4}^* \, P_L \, {\tilde
\chi}^0_i \;\;\;.\;\;\;
\end{eqnarray}
We also rewrite the two Higgs vevs in terms of the (tree level) $M_W$ and $M_Z$, using the notation~\cite{GH1986}: $M_W^2 = g^2 \left( v_1^2 + v_2^2 \right)/2 $ and  $\tan \beta \equiv v_2 / v_1$. We end up with
\begin{eqnarray}
{\cal L}_{\rm int } &\supset& - \frac{1}{M_P} \, {\bar \psi}_\mu \, P_L \Big[ \sqrt{2} \, M_W \left( \cos \beta \, U_{j2}^* \, W^{\mu +} {\tilde \chi}_j^c + \sin \beta \, V_{j2}^* \, W^{\mu -} {\tilde \chi}_j  \right) \nonumber\\
&&\quad\quad\quad\quad\quad\quad\quad\quad\quad\quad\quad\quad
+ M_Z \left( \cos \beta \, N_{i3}^* - \sin \beta \, N_{i4}^* \right)  Z^\mu \, {\tilde \chi}^0_i \Big] + {\rm h. c.}
\end{eqnarray}

The remaining interactions between the gravitino  and MSSM fields
coming from (\ref{eq:grav int L}), can be found in
\cite{Pradlerthesis} using MSSM gauge eigenstates in the absence of
electroweak symmetry breaking. In Appendix \ref{app:vertices-all},
we rewrite the gravitino-MSSM interactions in terms of the MSSM mass
eigenstates, including the effects of electroweak symmetry breaking
in the rotation matrices (between gauge and mass eigenstates).

The couplings of the gravitino at energies much greater than its
mass can be more easily written in terms of an effective interaction
between matter and the goldstino field \cite{Fayet}. The situation
is analogous to what happens for spontaneously broken gauge
theories, for which the couplings of the longitudinal polarization
of massive gauge bosons are determined at high energies by those of
the goldstone bosons that are eliminated in the unitary gauge (this
is known as the equivalence theorem). Analogously, the gravitino is
coupled at the quadratic level with the goldstino field. In the
super-Higgs mechanism, the goldstino is absorbed into a redefined
gravitino field (or, equivalently, it is set to zero in the unitary
gauge). At energies greater than the gravitino mass, the
longitudinal gravitino component is more strongly coupled to matter
than the transverse modes, and the couplings are determined by those
of the absorbed goldstino field (for a recent general study of the
phenomenology of a strongly coupled glodstino, see
\cite{Antoniadis:2010hs}).

We can see this from the polarization tensor \cite{VanNieuwenhuizen:1981ae}
\begin{eqnarray}
\Pi_{\mu \nu} &\equiv&  \sum_{r=\pm\frac{1}{2},\pm\frac{3}{2}} \, {\psi}_\mu^{(r)} \, {\bar \psi}_\nu^{(r)} \nonumber\\
&=& -(\slashed{p}+m_{3/2})\left(  g_{\mu\nu}-\frac{p_{\mu}p_{\nu}}{m_{3/2}^{2}}\right)  -\frac{1}{3}\left(  \gamma_{\mu}+\frac{p_{\mu}}{m_{3/2}}\right)  (\slashed{p}-m_{3/2})\left(  \gamma_{\nu}+\frac{p_{\nu}
}{m_{3/2}}\right)  \nonumber\\
&=& \frac{2}{3} \, \frac{p_\mu \, p_\nu \, \slashed{p}}{m_{3/2}^2}+ {\rm O } \left( \frac{1}{m_{3/2}} \right)
+ \cdots
\label{sum-grav-pol}
\end{eqnarray}
where $p$ and $m_{3/2}$ are the gravitino momentum and mass, respectively. In the last expression, we have written the leading term in the polarization tensor in a $1/m_{3/2}$ expansion.
This term, which
comes from the longitudinal gravitino polarization, dominates at energies greater than the gravitino mass. Since $\sum_{r=\pm\frac{1}{2}} \, {\chi}^{(r)} \, {\bar \chi}^{(r)} = \slashed{p} + m_{3/2} \approx \slashed{p}$, we can effectively replace $\psi_\mu \rightarrow \sqrt{\frac{2}{3}} \, \frac{\partial_\mu \chi}{m_{3/2}}$ (up to an irrelevant phase) in this high energy regime. Therefore (after an integration by parts)
\begin{equation}
{\cal L}_{\rm int} = - \frac{i}{\sqrt{2} \, M_P} \, {\bar \psi}_\mu \, S_{\rm MSSM}^\mu + {\rm h. \, c.} \;\;\rightarrow\;\;
{\cal L}_{\rm int, \, eff} = \frac{i}{\sqrt{3} \, m_{3/2} \, M_P} \,  {\bar \chi} \, \partial_\mu \, S_{\rm MSSM}^\mu + {\rm h. \, c.}
\label{eq:equivalence}
\end{equation}

We can actually simplify this expression  further, and obtain an
effective interaction Lagrangian in non-derivative form. To see
this, consider the infinitesimal variation of the MSSM Lagrangian
${\cal L}_{\rm MSSM} = {\cal L}_{\rm susy} + {\cal L}_{\rm soft} $
under an arbitrary infinitesimal variation of the MSSM fields. Since
only MSSM fields or their first derivatives enter in the Lagrangian,
one has
\begin{equation}
\delta {\cal L}_{\rm MSSM} = \frac{\partial {\cal L}_{\rm MSSM}}{\partial \Phi_i} \, \delta \Phi_i +
\frac{\partial {\cal L}_{\rm MSSM}}{\partial \left( \partial_\mu \Phi_i \right)} \, \delta \partial_\mu \Phi_i .
\end{equation}
One can then immediately rewrite this expression as
\begin{equation}
\delta {\cal L}_{\rm MSSM} = \left[ \frac{\partial {\cal L}_{\rm MSSM}}{\partial \Phi_i} - \partial_\mu \,
\frac{\partial {\cal L}_{\rm MSSM}}{\partial \left( \partial_\mu \Phi_i \right)} \right]  \delta \Phi_i
+ \partial_\mu \left( \frac{\partial {\cal L}_{\rm MSSM}}{\partial \left( \partial_\mu \Phi_i \right)} \, \delta \Phi_i
\right) .
\label{eq:proof1}
\end{equation}
Let us now specify the infinitesimal variations $\delta \Phi_i$ to be the variations of the MSSM fields under a supersymmetry transformation. Using eq. (\ref{eq:supercurrent}), and the fact that ${\cal L}_{\rm soft}$ does not contain first derivatives of fields, we have
\begin{equation}
\partial_\mu \left( \frac{\partial {\cal L}_{\rm MSSM}}{\partial \left( \partial_\mu \Phi_i \right)} \, \delta \Phi_i \right) = \partial_\mu \left( \frac{\partial {\cal L}_{\rm susy}}{\partial \left( \partial_\mu \Phi_i \right)} \, \delta \Phi_i \right) = \partial_\mu \left[ S_{\rm MSSM}^\mu + K^\mu \right]
\label{eq:proof2}
\end{equation}
where we recall that $\partial_\mu \, K^\mu$ is the variation of ${\cal L}_{\rm susy}$ under an infinitesimal supersymmetry transformation. Therefore
\begin{equation}
\partial_\mu K^\mu = \delta {\cal L}_{\rm susy} = \delta {\cal L}_{\rm MSSM} - \delta {\cal L}_{\rm soft} =
\delta {\cal L}_{\rm MSSM} - \frac{\partial {\cal L}_{\rm soft}}{\partial \Phi_i} \, \delta \Phi_i .
\end{equation}
Inserting this equation in eq. (\ref{eq:proof2}), and the resulting expression in eq. (\ref{eq:proof1}), we obtain
\begin{equation}
\delta {\cal L}_{\rm MSSM} = \left[ \frac{\partial {\cal L}_{\rm MSSM}}{\partial \Phi_i} - \partial_\mu \,
\frac{\partial {\cal L}_{\rm MSSM}}{\partial \left( \partial_\mu \Phi_i \right)} \right]  \delta \Phi_i + \partial_\mu \, S_{\rm MSSM}^\mu + \delta {\cal L}_{\rm MSSM} - \frac{\partial {\cal L}_{\rm soft}}{\partial \Phi_i} \, \delta \Phi_i ,
\end{equation}
or
\begin{equation}
\partial_\mu \, S_{\rm MSSM}^\mu = \left\{ \frac{\partial {\cal L}_{\rm soft}}{\partial \Phi_i} - \left[ \frac{\partial {\cal L}_{\rm MSSM}}{\partial \Phi_i} - \partial_\mu \,
\frac{\partial {\cal L}_{\rm MSSM}}{\partial \left( \partial_\mu \Phi_i \right)} \right] \right\} \delta \Phi_i
\end{equation}

Inserting this expression in eq. (\ref{eq:equivalence}), we rewrite
the interaction Lagrangian between the MSSM and the light gravitino
as
\begin{equation}
{\cal L}_{\rm int, \, eff} = \frac{i}{\sqrt{3} \, m_{3/2} \, M_P} \,  {\bar \chi} \, \left\{ \frac{\partial {\cal L}_{\rm soft}}{\partial \Phi_i} - \left[ \frac{\partial {\cal L}_{\rm MSSM}}{\partial \Phi_i} - \partial_\mu \,
\frac{\partial {\cal L}_{\rm MSSM}}{\partial \left( \partial_\mu \Phi_i \right)} \right]  \right\}  \delta \Phi_i  + {\rm h. \, c.}
\label{eq:non-der-beforeproof}
\end{equation}

As we prove in Appendix \ref{app:equivalence}, the part in square parenthesis does not contribute to the amplitudes of physical processes having one light gravitino in the initial or final state (in short, one can take the on shell expression for $\partial_\mu \, S_{\rm MSSM}^\mu$, since the term in square parenthesis vanishes on shell; notice that the procedure just outlined provides the on-shell expression of $\partial_\mu \, S_{\rm MSSM}^\mu$ without the need to explicitly work out the equations of motion of the fields entering in the supercurrent). Namely:
\begin{equation}
{\cal L}_{\rm int, \, eff} = \frac{i}{\sqrt{3} \, m_{3/2} \, M_P} \,  {\bar \chi} \,  \frac{\partial {\cal L}_{\rm soft}}{\partial \Phi_i}   \delta \Phi_i + {\rm h. \, c.}
\label{eq:non-der}
\end{equation}
This is the effective theory for the MSSM-light gravitino interaction in non-derivative form. To get an explicit expression, we recall the MSSM superpotential and soft supersymmetry breaking Lagrangian:
\begin{eqnarray}
W &=&  h_u H_2 Q u^c +  h_d H_1 Q d^c + h_e H_1 L e^c + \mu H_2 H_1  \\
{\cal L}_{\rm soft}  &= & -\left(\frac{1}{2} M_\alpha \lambda^\alpha
\lambda^\alpha + h.c. \right)- m_{ij}^2 {\phi^i}^* \phi^j \\
\nonumber && -\left(A_u h_u H_2 Q u^c + A_d h_d H_1 Q d^c +A_e h_e
H_1 L e^c + B \mu H_2 H_1 + h.c.\right) \label{lsoft}
\end{eqnarray}
where generation indices on the matter fields have been suppressed.
From this, we find
\begin{eqnarray}
{\cal L}_{\rm int, \, eff} &=& \frac{i \, m_{ij}^2}{\sqrt{3} \, M_P
\, m_{3/2}} \left( {\bar \chi} \, \chi_L^i \, \phi^{* j} - {\bar
\chi}_L^i \, \chi \, \phi^j \right) + \frac{i}{\sqrt{3} \, M_P \,
m_{3/2}} \left[  A_j W_{j,i}
\, {\bar \chi} \, \chi_L^i - \left( A_j W_{j,i} \right)^* \, {\bar \chi}_L^i \, \chi \right] \nonumber\\
&& - \frac{M_\alpha}{4 \, \sqrt{6} \, M_P \, m_{3/2}} \, F_{\mu \nu}^{(\alpha) a} \, {\bar \chi} \left[ \gamma^\mu ,\, \gamma^\nu \right] \, \lambda^{(\alpha) a} - \frac{i \, g_\alpha \, M_\alpha}{\sqrt{6} \, M_P \, m_{3/2}} \left( \phi^{*i} \, T_{ij}^a \, \phi^j \right) {\bar \chi} \, \gamma^5 \, \lambda^{(\alpha) a}
\label{L-int-eff}
\end{eqnarray}
where $\left( \phi_i ,\, \chi_L^i \right)$ are the MSSM chiral multiplets, $m_{ij}^2$ are the (low energy)
scalar mass$^2$ terms. In the second term, $A_j = A_u, A_d, A_e$, and $B$ for $j = 1 - 4$,
and $W_j$ refers to the respective term in the superpotential.
The indices $j$ and $i$ are both summed and the latter runs over the chiral fields in each $W_j$.

\section{Computation of $W^+ \, W^- \, \rightarrow \, $ gravitino $+$ neutralino}
\label{sect:comp}

We now compute the cross section for the scattering of unpolarized  W pairs
\begin{equation}
W^+ \left( k \right) + W^- \left( k' \right) \rightarrow {\tilde G}
\left( p \right) + {\tilde \chi}_i^{0} \left( q \right) ,
\end{equation}
which was studied in \cite{Ferrantelli:2007bx}. There are five diagrams contributing to this process, which we show in Figure \ref{fig:wwdiagrams}. The contribution from each diagram to the amplitude of the process can be found in Appendix \ref{app:amplitudes}.

\begin{figure}[ht!]
\begin{center}
\epsfig{file=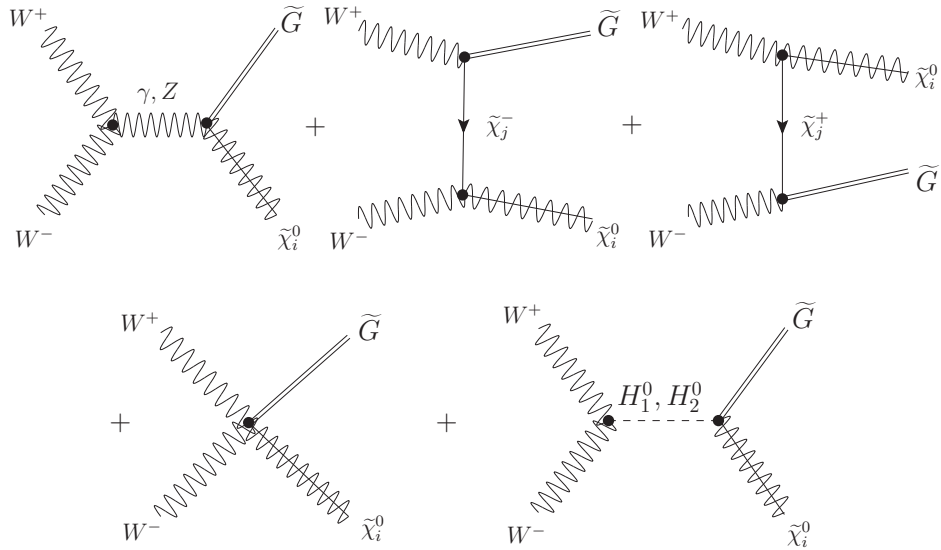,width=0.80\textwidth}
\end{center}
\caption{
Diagrams contributing to $W^+ \, W^- \, \rightarrow \, $ gravitino $+$ neutralino. }
\label{fig:wwdiagrams}
\end{figure}

We denote by ${\cal M}_i$ the matrix element for each neutralino mass eigenstate ${\tilde \chi}_i^0$ produced in the scattering. We computed the unpolarized squared matrix element
\begin{equation}
\overline{\vert {\cal M} \vert_i^2} \equiv \frac{1}{9} \, \sum_{\rm spins ,\, polarizations} {\cal M}_i \, {\cal M}_i^\dagger
\end{equation}
with the aid of the Mathematica package FeynCalc \cite{Mertig:1990an} (note that the index $i$ is not summed over). The resulting exact expressions are too long to be reported here.

The matrix elements are dimensionless, and since all gravitino
vertices are proportional to $1/M_P$, all of the squares of the
matrix elements are proportional to $1/M_P^2$. The sum over the
gravitino polarizations produces a term proportional to
$1/m_{3/2}^2$, cf. eq. (\ref{sum-grav-pol}). As a consequence, the
strength of the gravitino interactions increase with decreasing
$m_{3/2}$, and one can use accelerator constraints to set a lower
limit on the gravitino mass as discussed earlier. In  Ref.
\cite{Ferrantelli:2007bx}, it was claimed that the term proportional
to $1/m_{3/2}^2$ results in  a contribution $\overline{\vert {\cal
M} \vert_i^2} = \frac{X}{M_P^2 \, m_{3/2}^2}$ to the unpolarized
squared matrix elements, with a coefficient $X$ that does not vanish
in the limit of exact supersymmetry. More precisely, $X$ was found
to be proportional to the sum of a product of two Mandelstam
variables. This is quite different from the case of gravitino
production from gluon fusion, where the analog of $X$ is
proportional to the square of the gluino mass and hence vanishes in
the limit of exact SUSY. The discrepancy was attributed in
\cite{Ferrantelli:2007bx} to the breaking of the electroweak gauge
symmetry.

Although the complete result for the amplitude is too long to be reported here, we show in Fig. \ref{fig:compare} the resulting cross section for a specific choice of parameters (see below for details). Moreover, we analytically study, and present, the result in two relevant limits. Several combinations of the physical masses and rotation parameters
\begin{equation}
M_{H_1^0}^2,\,M_{H_2^0}^2,\,\beta,\,\alpha
,\,M_{{\tilde \chi}_j},\,U_{1j},\,U_{2j},\,V_{1j},\,V_{2j}
,\,M_{{\tilde \chi}_i^0},\,\,N_{i1},\,N_{i2},\,N_{i3},\,N_{i4}
\label{parameters-needed}
\end{equation}
appear in our (very lengthy) exact results for $\overline{\vert {\cal M} \vert_i^2}$.
$\alpha$ is the mixing angle in the scalar Higgs mass matrix, and the other quantities have been
defined above. As a consequence,
 the behavior of the exact expressions in the limit of exact supersymmetry is not manifest.
 The supersymmetry breaking parameters in the MSSM that are relevant for this computation are the soft masses $M_1,\, M_2,\,m_{H_u}^2,\,m_{H_d}^2,\,B$ introduced in eq. (\ref{lsoft}). To study the supersymmetric limit of  $\overline{\vert {\cal M} \vert_i^2}$, we assume that the soft SUSY masses are of the same order of magnitude, which we denote as $m_{\rm susy}$:
\begin{equation}
M_1 \sim M_2 \sim \vert m_{H_u} \vert \sim \vert m_{H_d} \vert \sim \vert B \vert = {\rm O } \left( m_{\rm susy} \right) \;\;\;,\;\;\; m_{3/2} \ll m_{\rm susy} \ll M_W
\label{msusy}
\end{equation}
The limit $m_{3/2} \ll m_{\rm susy}$ is the phenomenological relevant limit for setting accelerator bounds; the limit $m_{\rm susy} \ll M_W$, although not physically realized in Nature, is the appropriate assumption to analytically study the claim of \cite{Ferrantelli:2007bx}.

In the limit $m_{3/2} \ll m_{\rm susy} \ll M_W$, our averaged amplitudes can be formally written in the form~\footnote{We stress that all the terms of the sum (\ref{sum-grav-pol}) are included in our exact expressions from which the expansion (\ref{exp-amplitudes}) is performed.}
\begin{eqnarray}
&&\overline{\vert {\cal M} \vert_i^2} = \frac{\sum_{i,j =0} X_{ij} m_{\rm susy}^i m_{3/2}^j}
{M_P^2 \, m_{3/2}^2}
\label{exp-amplitudes}
\end{eqnarray}
where the coefficients of the expansions, $X_{ij}$ (some of which
may vanish) are independent of $m_{\rm susy}$ and $m_{3/2}$. Note
that the coefficients will be different for the different outgoing
neutralinos.

The only other relevant input parameter besides the soft masses (\ref{msusy}) is the $\mu$ parameter of the Higgs potential. The minimization of the Higgs potential leads to two equations
which allow one to solve for the two expectations values or equivalently,
$M_Z = (g^2 + g'^2) (v_1^2 + v_2^2)/2$ and $\tan \beta$.  Instead, it is common
to specify $M_Z$ and $\tan \beta$, in which case it is possible to solve for
the Higgs mass mixing parameter, $\mu$, and $B$,
\begin{eqnarray}
\mu^2 & = & \frac{m_{H_d}^2 - m_{H_u}^2 \tan^2 \beta + \frac{1}{2} M_Z^2 (1 -
\tan^2 \beta) + \Delta_\mu^{(1)}}{\tan^2 \beta - 1 + \Delta_\mu^{(2)}}
\nonumber \\
B \mu  & = & {1 \over 2} (m_{H_d}^2  + m_{H_u}^2 + 2 \mu^2) \sin 2
\beta + \Delta_B , \label{susylimit}
\end{eqnarray}
where $\Delta_B$ and $\Delta_\mu^{(1,2)}$ are one-loop corrections to $\mu$ and $B\mu$,
but will be ignored in our analytic expansions
as we are restricting our calculation of the cross section to tree level.
In the supersymmetric limit $\mu, B \to 0$ and $\tan \beta \to 1$ (see e.g. \cite{GH1986}).
From these expressions, we can express the parameters (\ref{parameters-needed}) as expansion series in $m_{\rm susy} / M_W$.~\footnote{The resulting expressions for (\ref{parameters-needed}) are lengthy, and we do not report them here.} Finally, we insert these expressions into $\overline{\vert {\cal M} \vert_i^2}$, and we find that the result
\begin{equation}
X_{00} = X_{01} = X_{10} = 0 \;\;\;{\rm in \; eq. } \; (\ref{exp-amplitudes})
\end{equation}
is indeed recovered for all $i=1,2,3,4$ (this explicitly shows that
the numerator of (\ref{exp-amplitudes}) vanishes for exact
supersymmetry, in contrast to what claimed in
\cite{Ferrantelli:2007bx}). The remaining terms are in general
nonvanishing. In the limit $m_{3/2} \ll m_{\rm susy} \ll M_W$, the
unpolarized squared matrix elements are dominated by the term
proportional to $X_{20}$ in eq. (\ref{exp-amplitudes}). Since
$X_{20}$ has mass dimension $2$, and since the external momenta can
be expressed in terms of Mandelstam  variables, we formally have
$X_{20} = d_0 \times {\rm momenta}^2 + d_2 \times \, M_W^2$, where
$d_{0,2}$ are combinations of dimensionless quantities (in practice,
only numerical factors and the gauge group charges). The full
expressions for $X_{20}$ are still too lengthy to be written here.
However, the terms proportional to $d_0$, dominate in the high energy limit,
and in this case we can write out analytic expressions for the
unpolarized squared matrix elements. For  $m_{3/2} \ll m_{\rm susy}
\ll M_W \ll \sqrt{\vert t \vert} ,\, \sqrt{s} $, we find
\begin{eqnarray}
\label{final-result}
\overline{\vert {\cal M} \vert_1^2} &\simeq& \frac{e^2 s \, \, M_2^2}{27 \, \sin^2 \theta_w \, M_P^2 \,
m_{3/2}^2}\\
\overline{\vert {\cal M} \vert_2^2} &\simeq& \frac{e^2 \, s \, \left[ M_1 \left( M_1 + 2 \, M_2 \right) \, \left( 1 + 2 \, \frac{t}{s} \right)^2  + M_2^2 \left( 17 + 36 \, \frac{t}{s} + 36 \, \frac{t^2}{s^2} \right) \right]}{108 \, M_P^2 \, m_{3/2}^2} \nonumber\\
\overline{\vert {\cal M} \vert_3^2} \simeq \overline{\vert {\cal M} \vert_4^2} &\simeq&
\frac{e^2 \, s}{216 \, M_P^2 \, m_{3/2}^2} \, \times \nonumber\\
&&\!\!\!\!\!\!\!\!\!\!\!\!\!\!\!\!\!\!\!\!\!\!\!
 \left\{ M_1 \left( \tan^2 \theta_w \, M_1 - 2 \, M_2 \right) \left( 1 + 2 \, \frac{t}{s} \right)^2 + \frac{M_2^2}{\sin^2 \theta_w } \left[ 4 + \left( 17 + 36 \, \frac{t}{s} + 36 \, \frac{t^2}{s^2} \right) \cos^2 \theta_w \right] \right\} \nonumber
\end{eqnarray}
In this limit, the masses of ${\tilde \chi}_1^{0}$ and ${\tilde
\chi}_2^{0}$ go to 0 (as $m_{\rm susy} \to 0$). These states are a
symmetric combination of the Higgsinos and the photino respectively.
The masses of $\chi_3$ and $\chi_4$ both approach $M_Z$, and these
are mixtures of the zino and an anti-symmetric and symmetric
combination of the Higgsinos.

We also computed the scattering $W^+ \, W^- \,
\rightarrow \, $ goldstino $+$ neutralino using the effective theory
(\ref{L-int-eff}), and we precisely recovered the expressions
(\ref{final-result}) in the high energy
limit. We note that the first and last term in
(\ref{L-int-eff}) also give quadratic goldstino-neutralino
interactions, proportional to the two Higgs vevs. Such terms are
included in the computation as mass insertions.

It is also interesting to study the exact results in the limit of $M_W \ll m_{\rm susy}$, since this is the more phenomenologically relevant one. Repeating the same exercise discussed above,
we find in the limit $m_{3/2} \ll M_W \ll m_{\rm susy} \ll \sqrt{\vert t \vert} ,\, \sqrt{s}$
\begin{eqnarray}
\label{final-result2}
\overline{\vert {\cal M} \vert_1^2} &\simeq& \frac{e^2  \, M_1^2}{108 \, \cos^2 \theta_w \, M_P^2 \,
m_{3/2}^2} \, \left[ \frac{\left( s + 2 t \right)^2}{s} + s \, \cos^2 \left( 2 \beta \right) \right]
\nonumber\\
\overline{\vert {\cal M} \vert_2^2} &\simeq& \frac{e^2  \, M_2^2}{108 \, \sin^2 \theta_w \, M_P^2 \,
m_{3/2}^2} \, \left[ \frac{9 \left( s + 2 t \right)^2}{s} + 8 \, s + s \, \cos^2 \left( 2 \beta \right)
\right]  \nonumber\\
\overline{\vert {\cal M} \vert_3^2} &\simeq& \overline{\vert {\cal M} \vert_4^2} \simeq
\frac{e^2  \, M_2^2}{27 \, \sin^2 \theta_w \, M_P^2 \,  m_{3/2}^2} \, s
\end{eqnarray}
As one would expect, we again find that the numerators vanish for $m_{\rm susy} \rightarrow 0$ in this limit. Now, the four neutralinos have masses which approach $M_1$, $M_2$,
$\mu$ and $\mu$ respectively and are effectively the bino, wino, and
antisymmetric and symmetric Higgsinos. As we did for eqs. (\ref{final-result}),
we also reproduced the results 
(\ref{final-result2}) 
using the effective goldstino
Lagrangian (\ref{L-int-eff}).

From both (\ref{final-result}) and (\ref{final-result2}) we see that the square amplitude grows linearly with the Mandelstam variables. The resulting cross section is therefore constant at $s^{1/2}$ much greater than the masses of the particles involved in the scattering. This is in contrast with the $\sigma \propto s$ dependence claimed in \cite{Ferrantelli:2007bx}.

For illustrative purposes, we show the cross section for a specific
choice of parameters. We choose to work in the context of no-scale
supergravity \cite{noscale} characterized by the K\"ahler potential
\begin{eqnarray}
G &=& \frac{K}{M_P^2} + F \left( \phi^i \right) + F^\dagger \left( \phi_i^* \right) \nonumber\\
K &=& - 3 \, M_P^2 \, {\rm ln } \left[ \frac{z + z^*}{M_P}  - \frac{\phi^i \, \phi_i^*}{3 \, M_P^2} \right]
\label{kahler}
\end{eqnarray}
where for simplicity we consider only one hidden sector complex
field, $z$. The scalar potential takes a globally supersymmetric
form \beq V = e^{G - \frac{1}{3}K} \left| W_{,i} \right|^2 \eeq plus
$D$-terms. It is important to note here the absence of all of the
soft supersymmetry breaking masses.  That is, at the scale at which
supergravity is broken (which we assume to be greater than the grand
unified scale), $m_0^2 = A_0 = B_0$ = 0.  These terms will be
generated radiatively from the non-zero gaugino mass, which at the
supersymmetry breaking scale is given by 
\beq 
m_\lambda =
\frac{1}{2} \left| e^{G/2} \frac{G_{,z}}{G_{,zz^*}} (\ln Re
h)^*_{,z^*} \right| , 
\eeq 
where $h(z)$ is the gauge kinetic
function assumed to be diagonal in its gauge indices. For the
no-scale K\"ahler potential, one then finds that 
\beq 
m_\lambda =
\frac{1}{2} m_{3/2}^{1/3} \frac{h_{,z}}{Re h} .
\eeq 
For a suitable
choice of $h$ \cite{noscale2},  the gravitino mass can be made much
smaller than the gaugino mass.

Phenomenological models based on no-scale supergravity have been recently constructed
\cite{emo}, and we use two examples of low energy spectra based on that work.
In the first example, we choose a supersymmetry breaking scale of $M_P$,
and a universal gaugino mass $m_{1/2} = 600$ GeV. Recall that $m_0 = A_0 = B_0 = 0$.
The low energy spectra also depend on two couplings in the GUT scale superpotential
corresponding to the term cubic in the Higgs adjoint ($\lambda'$) and a mixing term between
the adjoint and the Higgs 5-plets ($\lambda$).  In this example, we take $\lambda = -0.06$ and
$\lambda' = 1$.
Because we are specifying $B_0$ at the input scale, we are not free to choose $\tan \beta$.
In this example, it is calculated to be $\tan \beta = 47.8$. When run to the weak scale,
this model has gaugino masses of $M_1 = 275 \, {\rm GeV} , M_2 = 534 \, {\rm GeV} $.
The soft Higgs masses are
$m_{H_u}^2 = -1024^2 \, {\rm GeV} \;\; m_{H_d}^2 = -615^2 \, {\rm GeV}$.
When loop corrections are included in calculating the low energy spectrum,
we find $\mu = 840$ GeV, and neutralino masses of $283, 550, 913$ and 919 GeV.
The gluino mass is 1510 GeV. The scalar Higgs masses are 119 and 734 GeV.
We have fixed the gravitino mass to
$m_{3/2} = 10^{-4} \, {\rm eV}$.

\begin{figure}[ht!]
\begin{center}
\epsfig{file=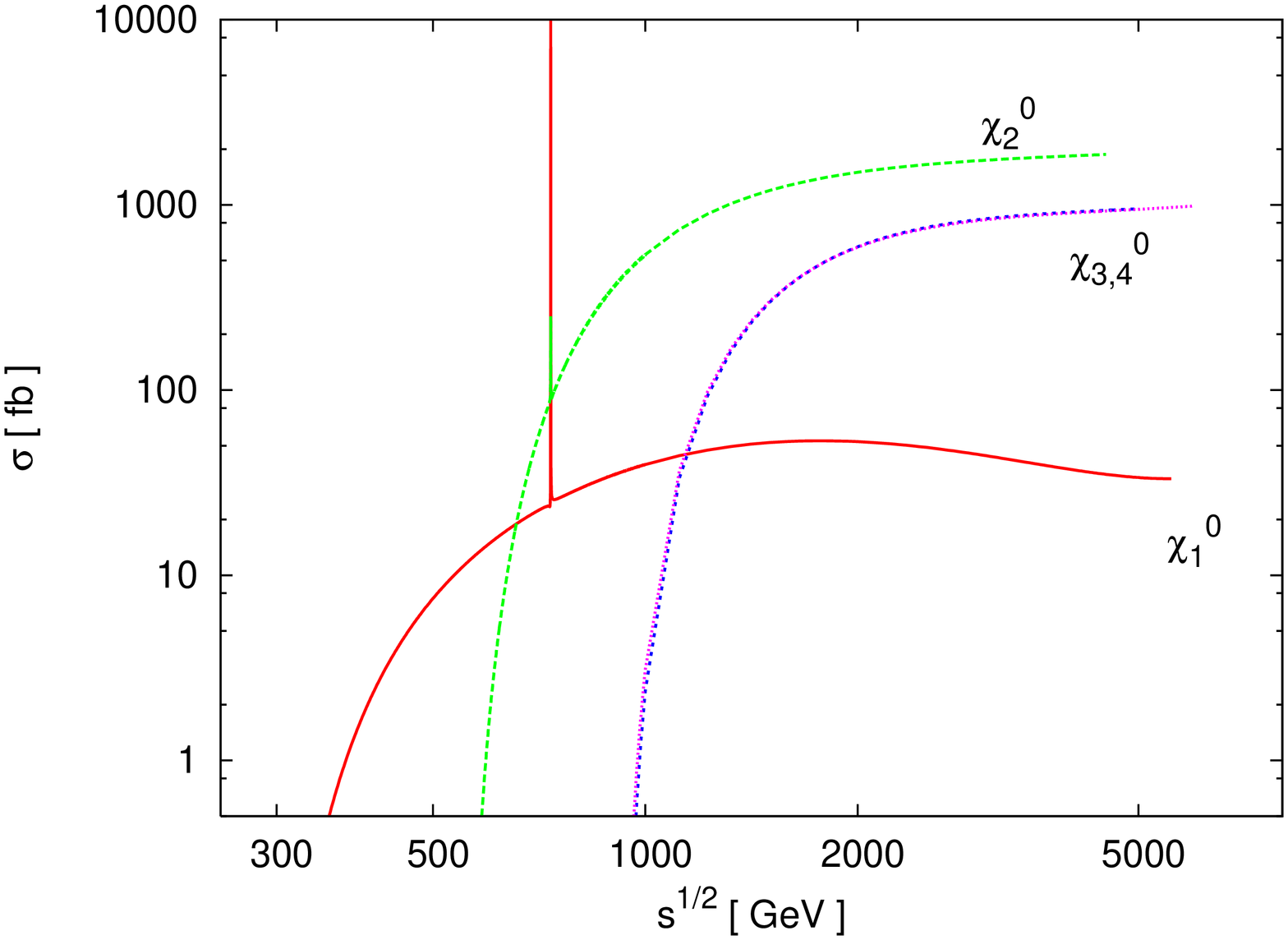,width=0.33\textwidth,angle=-90}
\epsfig{file=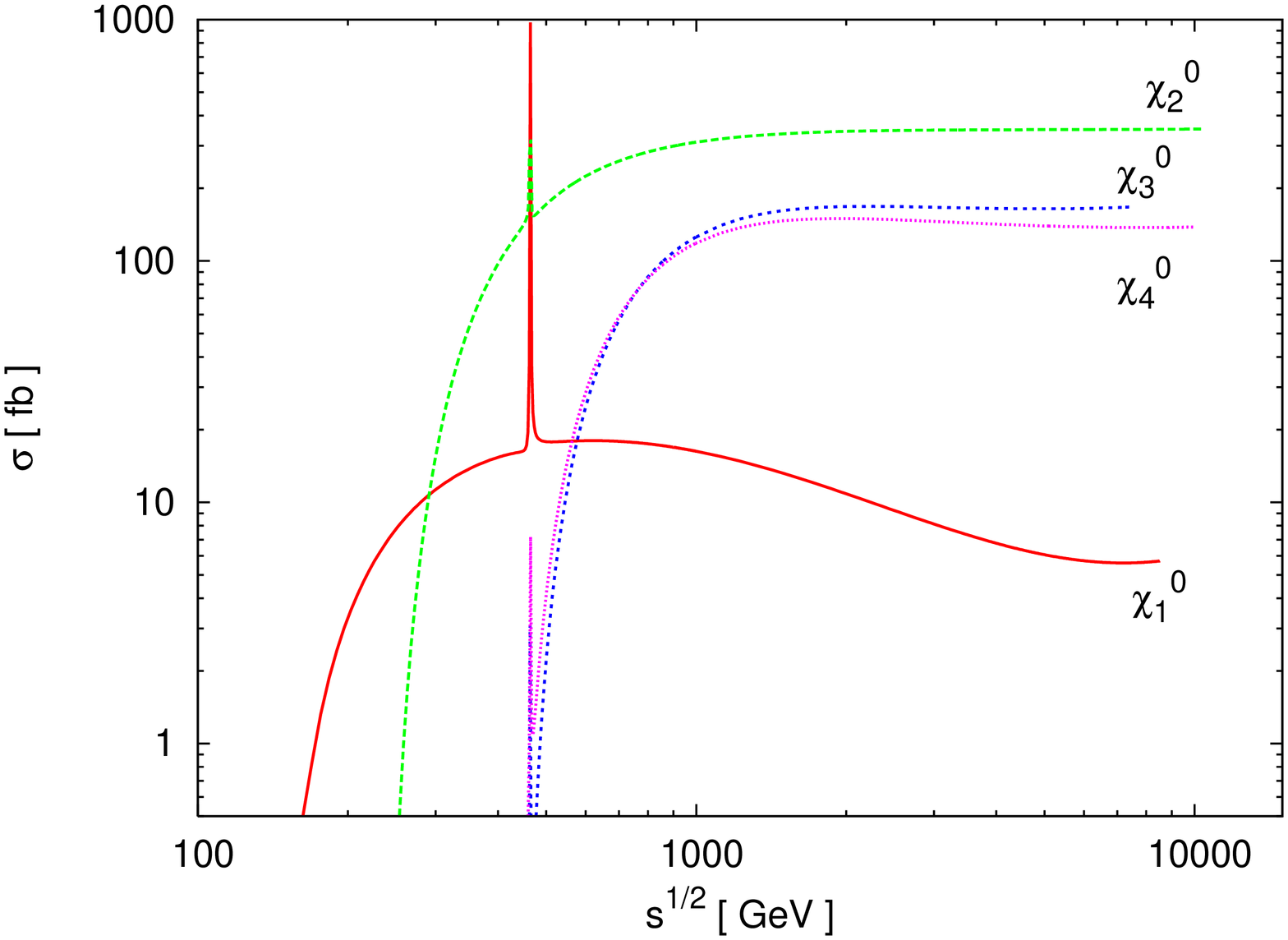,width=0.33\textwidth,angle=-90}
\end{center}
\caption{
Cross section for $W^+ \, W^- \, \rightarrow \, $ gravitino $+$ neutralino for the two sets of parameter choices specified in the main text. The gravitino mass has been fixed to $m_{3/2} = 10^{-4} \, {\rm eV}$. We recall that the cross section scales as $m_{3/2}^{-2}$.
 }
\label{fig:oursigma}
\end{figure}

We show in Figure \ref{fig:oursigma} the cross sections for the
production of a gravitino and each of the neutralino eigenstates.
These cross section are evaluated numerically from the exact square
amplitudes. We see that they indeed approach a constant value at
high $\sqrt{s}$. For this choice of parameters, the processes
producing the first two neutralinos have a resonance at $\sqrt{s} =
M_{H_1^0} $, corresponding to the heavy Higgs exchange process. The
resonance is narrow as compared to the range of $\sqrt{s}$ shown
here, and it is barely visible in the result for $\chi_{1}^0$ shown in
the Figure.

\begin{figure}[ht!]
\begin{center}
\epsfig{file=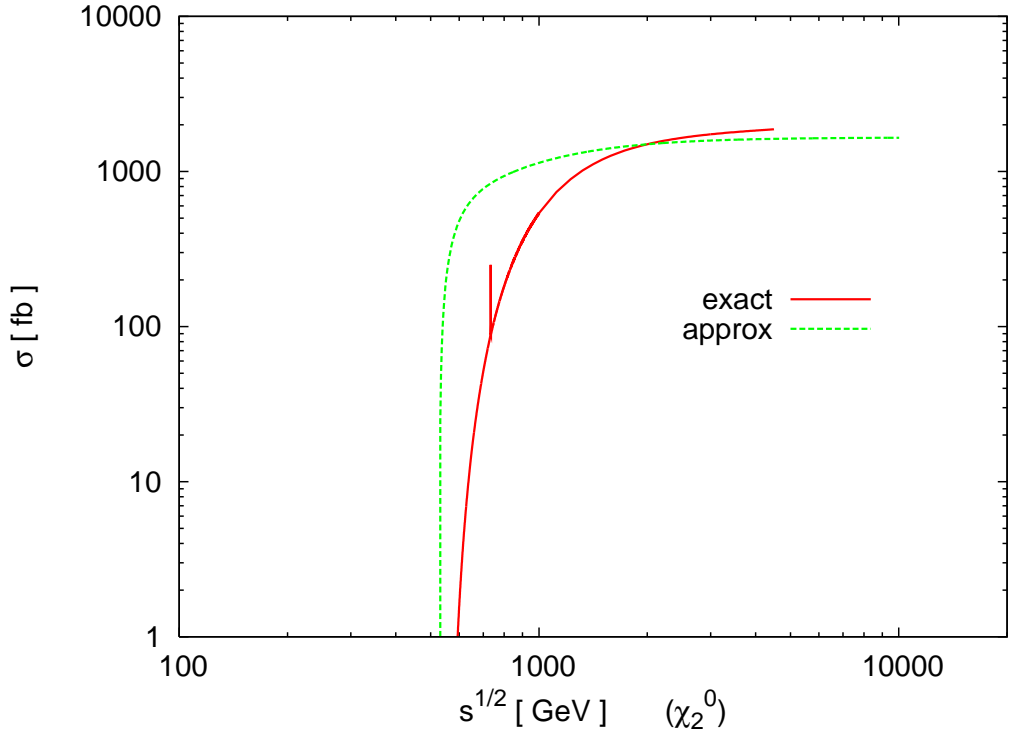,width=0.60\textwidth,angle=-90}
\end{center}
\caption{
Comparison between the exact cross section shown in the previous Figure (for the second neutralino eigenstate), and the one obtained from the approximated square amplitude (\ref{final-result2}).
 }
\label{fig:comparison}
\end{figure}

Figure \ref{fig:oursigma} shows the cross sections until $\sqrt{s}$
becomes too large, and our results are affected by numerical
inaccuracies. However, one can verify that the analytic
approximations written above are in excellent agreement with the
exact expressions in the high energy limit. In Fig.
\ref{fig:comparison}, we compare the exact results with the
approximations (\ref{final-result2}), for the illustrative case
shown in the left panel of Fig. \ref{fig:oursigma} where the soft
supersymmetric masses are all sufficiently greater (in magnitude)
than $m_W$. In Figure \ref{fig:comparison}, we present the
comparison for the process producing the second neutralino
eigenstate (which is the dominant one for this choice of
parameters). An equally excellent agreement is also found for the
processes producing the other three neutralinos.

To see the dependence on our particular choice of low energy spectrum, we show in the right panel of Fig. \ref{fig:oursigma} a second example. In this case,
we choose a supersymmetry breaking scale of $3 \times 10^{16}$ GeV,
and a universal gaugino mass $m_{1/2} = 300$ GeV,
with $\lambda = 0.2$ and $\lambda' = 1$.
We find $\tan \beta = 11.2$ and gaugino masses of $M_1 = 123 \, {\rm GeV} , M_2 = 239 \, {\rm GeV} $
The soft Higgs masses are
$m_{H_u}^2 = -477^2 \, {\rm GeV} \;\; m_{H_d}^2 = 189^2 \, {\rm GeV}$,
 $\mu = 413$ GeV, and neutralino masses of $123, 231, 426$ and 442 GeV.
The gluino mass is 745 GeV. The scalar Higgs masses are 113 and 464 GeV.
Again, we have fixed the gravitino mass to
$m_{3/2} = 10^{-4} \, {\rm eV}$.

In our second example, the sparticle spectrum is somewhat lighter (by roughly a factor of 2).
As one can see, the qualitative behavior of the cross sections is similar to that
found in the left panel.  The heavy Higgs resonance however, is now much more
prominent.

\section{Summary}
\label{sect:conc}

It is quite possible that low energy realizations of supersymmetry
yields a spectrum with a gravitino LSP. In models with gauge mediated supersymmetry
breaking, as well as in no-scale supergravity models, the gravitino may in fact be very light
compared with the rest of the superpartner spectrum.  As we have discussed above, and shown
rigorously in Appendix B, the high energy interactions of a light gravitino
are dominated by its longitudinal component, or goldstino.  As a result,
the couplings of gravitinos to matter are proportional to $1/M_P^2 m_{3/2}^2$
making them readily accessible in accelerator searches.

One might expect that gravitino production cross sections in $p$ $p$ collisions
be dominated by quark and gluon fusion process, however, a recent calculation
claimed that due to effects associated with electroweak symmetry breaking,
$W$ boson fusion process would eventually come to dominate
the overall gravitino production cross section at high energy.
Here, we have shown this claim to be untrue.  We have calculated
the gravitino production cross section in both the high energy and supersymmetric
limits and found no enhancement due to electroweak symmetry breaking effects.

Although the full analytic expression for the cross section for gravitino
production through $W$ fusion is too lengthy to write out, we
were able to express the cross section in two limiting cases both at
high energy: the supersymmetric limit - where one sees explicitly the
fact that the matrix elements are proportional to the gaugino masses
(just as they are for the case of an unbroken gauge symmetry) and
in the (more physical) electroweak limit where again the matrix elements are
proportional to gaugino masses.

We have also worked out in detail the applicability and use the equivalence theorem.
In the high energy limit, the interactions of the gravitino
can be replaced with derivative interactions of the goldstino.
After an integration by parts, the goldstino is coupled to the divergence of the
supercurrent.  In Appendix \ref{app:equivalence}, we prove the equivalence theorem
and show that its validity does not require an unbroken gauge symmetry.
As a consequence, we are able to write down a relatively simple form
for the effective interaction Lagrangian, and verify that
the resulting cross section agrees with the original result.

\section*{Acknowledgments}
We would like to thank X. Cui and M. Voloshin for helpful
discussions. This work was supported in part by DOE grant
DE-FG02-94ER-40823 at the University of Minnesota.

\appendix

\section{Single gravitino-MSSM vertices with broken electroweak symmetry}
\label{app:vertices-all}

We write here all the interactions of a single on-shell gravitino
with MSSM fields (we also include the term (\ref{int-propto-v})
worked out in the main text).  Besides the relations already written
in Section \ref{sec:interactions}, we also use the Higgs
decomposition \cite{GH1986}
\begin{eqnarray}
\label{eq: higgs gauge-mass}
&& H^1_2 = H^{+} \cos \beta, \,\,
H^2_1 = H^{-} \sin \beta, \nonumber \\
&& H^1_1 = v_1 + \frac{1}{\sqrt{2}}\left(H^0_1 \cos \alpha - H^0_2 \sin \alpha + {i} H^0_3 \sin \beta \right),
\nonumber \\
&& H^2_2 = v_2 + \frac{1}{\sqrt{2}}\left(H^0_1 \sin \alpha + H^0_2 \cos \alpha + {i} H^0_3 \cos \beta \right),
\end{eqnarray}
the sfermion rotation between the mass eigensates ${\tilde f}_1 ,\, {\tilde f}_2$ and the interaction eigenstates ${\tilde f}_L ,\, {\tilde f}_R$~\footnote{We denote the superfields associated to SM l.h. fermions by
$\left( P_L \nu ,\, {\tilde \nu}_L \right)$, $\left( P_L e ,\, {\tilde e}_L \right)$, $\left( P_L u ,\, {\tilde u}_L \right)$, $\left( P_L d ,\, {\tilde d}_L \right)$, and the superfields associated to SM r.h. fermions by
$\left( P_L e^c ,\, {\tilde e}_R^* \right)$, $\left( P_L u^c ,\, {\tilde u}_R^* \right)$, $\left( P_L d^c ,\, {\tilde d}_R^* \right)$; family indices are understood.
}
\begin{equation}
\left( \begin{array}{c} {\tilde f}_1 \\ {\tilde f}_2 \end{array} \right) =
\left( \begin{array}{cc} c_f & - s_f^* \\ s_f  & c_f^* \end{array} \right)
\left( \begin{array}{c} {\tilde f}_L \\ {\tilde f}_R \end{array} \right)
\;\;\;,\;\;\; \vert c_f \vert^2 + \vert s_f \vert^2 = 1
\end{equation}
and the relations between the gaugino, and neutralino/chargino mass eigenstates \cite{GH1986}
\begin{eqnarray}
{\tilde B} &=& \left( N_{i1}^* \, P_L + N_{i1} \, P_R \right) {\tilde \chi}_i^0  \nonumber\\
{\tilde W}^3 &=& \left( N_{i2}^* \, P_L + N_{i2} \, P_R \right) {\tilde \chi}_i^0  \nonumber\\
\frac{{\tilde A}^1 - i {\tilde A}^2}{\sqrt{2}} &\equiv& {\tilde W} =
\left( V_{j1}^* \, P_L + U_{j1} \, P_R \right) {\tilde \chi}_j \nonumber\\
\Rightarrow \;\;
\frac{{\tilde A}^1 + i {\tilde A}^2}{\sqrt{2}} &\equiv& {\tilde W}^c =
\left( U_{j1}^* \, P_L + V_{j1} \, P_R \right) {\tilde \chi}_j^c
\label{eq:def-nuv}
\end{eqnarray}

We find
\begin{eqnarray}
{\cal L}_{\rm int} &=& {\cal L}_{W {\tilde \chi}} + {\cal L}_{\gamma {\tilde \chi}^0} +
{\cal L}_{Z {\tilde \chi}^0} + {\cal L}_{H {\tilde \chi}} + {\cal L}_{H^0 {\tilde \chi}^0} +
{\cal L}_{f {\tilde f}} + {\cal L}_{G {\tilde G}}  \\
&&+ {\cal L}_{W H^0 {\tilde \chi}} + {\cal L}_{W H {\tilde \chi}^0} + {\cal L}_{W W {\tilde \chi}^0}
+ {\cal L}_{W f {\tilde f}} + {\cal L}_{\gamma H {\tilde \chi}} +{\cal L}_{\gamma f {\tilde f}}
+ {\cal L}_{Z H {\tilde \chi}} + {\cal L}_{Z H^0 {\tilde \chi}^0} + {\cal L}_{Z f {\tilde f}} \nonumber\\
&&+ {\cal L}_{W \gamma {\tilde \chi}} + {\cal L}_{W Z {\tilde \chi}} + {\cal L}_{G G {\tilde G}} +
{\cal L}_{G f {\tilde f}} \nonumber
\label{eq:all-formal}
\end{eqnarray}
where
\begin{eqnarray}
{\cal L}_{W {\tilde \chi}} &=& - \frac{{\bar \psi}_\mu}{M_P} \, \Big[ \left( \sqrt{2} \, M_W \, \cos \beta \, U_{j2}^* \, \eta^{\mu \sigma}
+ \frac{i}{4} \, \left[ \gamma^\rho ,\, \gamma^\sigma \right] \gamma^\mu \, U_{j1}^* \partial_\rho^{(W)} \right) P_L \nonumber\\
&&\quad\quad\quad\quad
+ \left( \sqrt{2} \, M_W \, \sin \beta \, V_{j2} \, \eta^{\mu \sigma}
+ \frac{i}{4} \, \left[ \gamma^\rho ,\, \gamma^\sigma \right] \gamma^\mu \, V_{j1} \partial_\rho^{(W)} \right) P_R \Big] W_\sigma^+ \, {\tilde \chi}_j^c \nonumber\\
&& - \frac{{\bar \psi}_\mu}{M_P} \, \Big[ \left( \sqrt{2} \, M_W \, \cos \beta \, U_{j2} \, \eta^{\mu \sigma}
+ \frac{i}{4} \, \left[ \gamma^\rho ,\, \gamma^\sigma \right] \gamma^\mu \, U_{j1} \partial_\rho^{(W)} \right) P_R \nonumber\\
&&\quad\quad\quad\quad
+ \left( \sqrt{2} \, M_W \, \sin \beta \, V_{j2}^* \, \eta^{\mu \sigma}
+ \frac{i}{4} \, \left[ \gamma^\rho ,\, \gamma^\sigma \right] \gamma^\mu \, V_{j1}^* \partial_\rho^{(W)} \right) P_L \Big] W_\sigma^- \, {\tilde \chi}_j \nonumber\\
\end{eqnarray}
\begin{equation}
{\cal L}_{\gamma {\tilde \chi}^0} = - \frac{i}{4 \, M_P} \, {\bar \psi}_\mu \, \left[ \gamma^\rho ,\, \gamma^\sigma \right] \gamma^\mu \, \partial_\rho \, A_\sigma \left[ \left( \cos \theta_w \, N_{i1}^* + \sin \theta_w \, N_{i2}^* \right) P_L +
\left( \cos \theta_w \, N_{i1} + \sin \theta_w \, N_{i2} \right) P_R \right] {\tilde \chi}_i^0
\end{equation}
\begin{eqnarray}
{\cal L}_{Z {\tilde \chi}^0} &=& - \frac{1}{M_P} {\bar \psi}_\mu \Big\{
\big[ M_Z \left( \cos \beta \, N_{i3}^* - \sin \beta \, N_{i4}^* \right) \eta^{\mu \sigma} \nonumber\\
&&\quad\quad\quad\quad
+ \frac{i}{4}
 \, \left[ \gamma^\rho ,\, \gamma^\sigma \right] \gamma^\mu \, \left( \cos \theta_w \, N_{i2}^* - \sin \theta_w \, N_{i1}^* \right) \partial_\rho^{(Z)} \big] P_L + \big[ M_Z \left( \cos \beta \, N_{i3} - \sin \beta \, N_{i4} \right) \eta^{\mu \sigma} \nonumber\\
&&\quad\quad\quad\quad + \frac{i}{4}
 \, \left[ \gamma^\rho ,\, \gamma^\sigma \right] \gamma^\mu \, \left( \cos \theta_w \, N_{i2} - \sin \theta_w \, N_{i1} \right) \partial_\rho^{(Z)} \big] P_R
 \Big\} Z_\sigma \, {\tilde \chi}_i^{(0)} \nonumber\\
\end{eqnarray}
\begin{eqnarray}
{\cal L}_{H {\tilde \chi}} &=& - \frac{1}{M_P} \, {\bar \psi}_\mu \, \sqrt{2} \, i \, \partial^\mu H^+
\left( \sin \beta \, U_{j2}^* \, P_L - \cos \beta \, V_{j2} \, P_R \right) {\tilde \chi}_j^c + {\rm h. \, c.}
\end{eqnarray}
\begin{eqnarray}
{\cal L}_{H^0 {\tilde \chi}^0} &=& - \frac{1}{M_P} \, {\bar \psi}_\mu \big\{
i  \left[ \left( \cos \alpha \, N_{i3}^* + \sin \alpha \, N_{i4}^* \right) P_L
- \left( \cos \alpha \, N_{i3} + \sin \alpha \, N_{i4} \right) P_R \right]
\partial^\mu H_1^0 \, {\tilde \chi}_i^{0} \nonumber\\
&&\quad\quad\quad\quad
+i  \left[ \left( -\sin \alpha \, N_{i3}^* + \cos \alpha \, N_{i4}^* \right) P_L
- \left( - \sin \alpha \, N_{i3} + \cos \alpha \, N_{i4} \right) P_R \right]
\partial^\mu H_2^0 \, {\tilde \chi}_i^{0} \nonumber\\
&&\quad\quad\quad\quad
+\left[ \left( \sin \beta \, N_{i3}^* + \cos \beta \, N_{i4}^* \right) P_L
+ \left( \sin \beta \, N_{i3} + \cos \beta \, N_{i4} \right) P_R \right]
\partial^\mu H_3^0 \, {\tilde \chi}_i^{0} \Big\}
\end{eqnarray}
\begin{equation}
{\cal L}_{f {\tilde f}} = - \frac{1}{M_P} \, {\bar \psi}_\mu \, \sqrt{2} \, i \left[
\sum_{f=\nu,e,u,d} \left( c_f \, \partial^\mu {\tilde f}_1^*
+ s_f \, \partial^\mu {\tilde f}_2^*  \right) P_L \, f +
\sum_{f=e,u,d}   \left( - s_f \, \partial^\mu {\tilde f}_1
+ c_f \, \partial^\mu {\tilde f}_2  \right) P_L \, f^c
\right]  + {\rm h. \, c.}
\end{equation}
\begin{equation}
{\cal L}_{G {\tilde G}}  = - \frac{i}{4 \, M_P} \, {\bar \psi}_\mu \, \left[ \gamma^\rho ,\, \gamma^\sigma \right] \gamma^\mu \, \partial_\rho \, G_\sigma^a  \, {\tilde G}^a
\end{equation}
\begin{eqnarray}
{\cal L}_{W H^0 {\tilde \chi}} &=& - \frac{1}{M_P} \, {\bar \psi}_\mu \, \frac{g}{\sqrt{2}} \Big[
\left( \cos \alpha \, U_{j2}^* \, P_L + \sin \alpha \, V_{j2} \, P_R \right) H_1^0
+ \left( - \sin \alpha \, U_{j2}^* \, P_L + \cos \alpha \, V_{j2} \, P_R \right) H_2^0 \nonumber\\
&&\quad\quad\quad\quad\quad\quad
+ i \left( - \sin \beta \, U_{j2}^* \, P_L + \cos \beta \, V_{j2} \, P_R \right) H_3^0 \Big] W^{\mu +} \, {\tilde \chi}_j^c
+ {\rm h. \, c.}
\end{eqnarray}
\begin{equation}
{\cal L}_{W H {\tilde \chi}^0} = - \frac{1}{M_P} \, {\bar \psi}_\mu \, g
\left( \cos \beta \, N_{i4}^* \, P_L + \sin \beta \, N_{i3} \, P_R \right)
W^{\mu +} \, H^- \, {\tilde \chi}_i^0 + {\rm h. \, c.}
\end{equation}
\begin{equation}
{\cal L}_{W W {\tilde \chi}^0} =
\frac{g}{4 \, M_P} \, {\bar \psi}_\mu \, \left[ \gamma^\rho ,\, \gamma^\sigma \right] \gamma^\mu
\left( N_{i2}^* \, P_L + N_{i2} \, P_R \right) W_\rho^+ \, W_\sigma^- \, {\tilde \chi}_i^0
\end{equation}
\begin{eqnarray}
{\cal L}_{W f {\tilde f}} &=& - \frac{1}{M_P} \, {\bar \psi}_\mu \, g \, \Big\{
W^{\mu +} \left[
\left( c_\nu \, {\tilde \nu}_1^* + s_\nu \, {\tilde \nu}_2^* \right) P_L \, e +
\left( c_u \, {\tilde u}_1^* + s_u \, {\tilde u}_2^* \right) P_L \, d
\right] \nonumber\\
&&\quad\quad\quad\quad\quad
+ W^{\mu -} \left[
\left( c_e \, {\tilde e}_1^* + s_e \, {\tilde e}_2^* \right) P_L \, \nu +
\left( c_d \, {\tilde d}_1^* + s_d \, {\tilde d}_2^* \right) P_L \, u
\right] \Big\} + {\rm h. \, c.}
\end{eqnarray}
\begin{equation}
{\cal L}_{\gamma H {\tilde \chi}} = - \frac{1}{M_P} \, {\bar \psi}_\mu \, \sqrt{2} \, e
\left( - \sin \beta \, U_{j2}^* \, P_L + \cos \beta \, V_{j2} \, P_R \right)
A^\mu \, H^+ \, {\tilde \chi}_j^c + {\rm h. \, c.}
\end{equation}
\begin{equation}
{\cal L}_{\gamma f {\tilde f}} = - \frac{1}{M_P} \, {\bar \psi}_\mu \,
\sqrt{2} \, e \, A^\mu \sum_{f=e,u,d}
q_f \left[  \left( c_f \, {\tilde f}_1^* + s_f \, {\tilde f}_2^* \right) P_L \, f
- \left( - s_f \, {\tilde f}_1 + c_f \, {\tilde f}_2 \right)  P_L \, f^c \right] + {\rm h. \, c.}
\end{equation}
\begin{equation}
{\cal L}_{Z H {\tilde \chi}} = - \frac{1}{M_P} \, {\bar \psi}_\mu \, \frac{g \left( \cos^2 \theta_w - \sin^2 \theta_w \right) }{\sqrt{2} \, \cos \theta_w} \left( - \sin \beta \, U_{j2}^* \, P_L + \cos \beta \, V_{j2} \, P_R \right)
Z^\mu \, H^+ \, {\tilde \chi}_j^c + {\rm h. \, c.}
\end{equation}
\begin{eqnarray}
{\cal L}_{Z H^0 {\tilde \chi}^0} &=& - \frac{1}{M_P} \, {\bar \psi}_\mu \, \frac{g}{2 \, \cos \, \theta_w} \Big\{
\left[ \left( \cos \alpha \, N_{i3}^* - \sin \alpha \, N_{i4}^* \right) P_L + \left( \cos \alpha \, N_{i3} - \sin \alpha \, N_{i4} \right) P_R \right] Z^\mu \, H_1^0 \, {\tilde \chi}_i^0 \nonumber\\
&&\quad\quad\quad\quad\quad\quad\quad\quad -
\left[ \left( \sin \alpha \, N_{i3}^* + \cos \alpha \, N_{i4}^* \right) P_L + \left( \sin \alpha \, N_{i3} + \cos \alpha \, N_{i4} \right) P_R \right] Z^\mu \, H_2^0 \, {\tilde \chi}_i^0 \nonumber\\
&&\quad\quad\quad\quad\quad\quad\quad\quad -
i \left[ \left( \sin \beta \, N_{i3}^* - \cos \beta \, N_{i4}^* \right) P_L - \left( \sin \beta \, N_{i3} - \cos \beta \, N_{i4} \right) P_R \right] Z^\mu \, H_3^0 \, {\tilde \chi}_i^0 \Big\} \nonumber\\
\end{eqnarray}
\begin{equation}
{\cal L}_{Z f {\tilde f}} = - \frac{1}{M_P} \, {\bar \psi}_\mu \, \frac{\sqrt{2} \, g \, Z^\mu}{\cos \theta_w} \left[
\sum_{f=\nu,e,u,d} Z_L^f \left( c_f \, {\tilde f}_1^* + s_f \, {\tilde f}_2^* \right) P_L \, f -
\sum_{f=e,u,d} Z_R^f \left( - s_f \, {\tilde f}_1 + c_f \, {\tilde f}_2 \right)  P_L \, f^c \right] + {\rm h. \, c.}
\end{equation}
\begin{equation}
{\cal L}_{W \gamma {\tilde \chi}} = - \frac{g}{4 \, M_P} \, {\bar \psi}_\mu \, \left[ \gamma^\rho ,\, \gamma^\sigma \right] \gamma^\mu \, W_\rho^+ \, \sin \theta_w A_\sigma
\left( U_{j1}^* \, P_L + V_{j1} \, P_R \right) {\tilde \chi}_j^c + {\rm h. c.}
\end{equation}
\begin{equation}
{\cal L}_{W Z {\tilde \chi}} = - \frac{g}{4 \, M_P} \, {\bar \psi}_\mu \, \left[ \gamma^\rho ,\, \gamma^\sigma \right] \gamma^\mu \, W_\rho^+ \, \cos \theta_w Z_\sigma
\left( U_{j1}^* \, P_L + V_{j1} \, P_R \right) {\tilde \chi}_j^c + {\rm h. c.}
\end{equation}
\begin{equation}
{\cal L}_{G G {\tilde G}} = \frac{i \, g_s}{8 \, M_P} \, {\bar \psi}_\mu \, \left[ \gamma^\rho ,\, \gamma^\sigma \right] \gamma^\mu \, f^{abc} \, {\tilde G}^a \, G_\rho^b \, G_\sigma^c
\end{equation}
\begin{equation}
{\cal L}_{G f {\tilde f}} = - \frac{1}{M_P} \, {\bar \psi}_\mu \,
\frac{g_s}{\sqrt{2}} G^{a \mu} \sum_{f=u,d} \left[
\left( c_f \, \lambda^{a*} \, {\tilde f}_1^* +
s_f \, \lambda^{a*} \, {\tilde f}_2^* \right)_i \, P_L \, f_i
- \left( - s_f \, \lambda^{a} \, {\tilde f}_1 +
c_f \, \lambda^{a} \, {\tilde f}_2 \right)_i \, P_L \, f_i^c
\right] + {\rm h.c.}
\end{equation}

In the above expressions, $\partial^{(W)}$, and $\partial^{(Z)}$
denote a derivative acting only on the $W$, and $Z$ fields,
respectively; $q_f$ denotes the electric charge of the fermion $f$;
$Z_L^f = \frac{1}{2}, \, -\frac{1}{2} + \sin^2 \theta_w, \,
\frac{1}{2} - \frac{2}{3} \sin^2 \theta_w, \, -\frac{1}{2} +
\frac{1}{3} \sin^2 \theta_w$ for $\nu, e, u, d$, respectively;
$Z_R^f = \sin^2 \theta_w, \, -\frac{2}{3} \sin^2 \theta_w, \,
\frac{1}{3} \sin^2 \theta_w$ for $e, u, d$, respectively;
$\lambda^{a}$ are the Gell-Mann matrices. Moreover, in some of the
above expressions we have also used the identities
\begin{equation}
\left( {\bar \psi}_\mu \, P_L \, {\tilde \chi}_j \right)^\dagger = {\bar{\tilde \chi}}_j \, P_R \, \psi_\mu =
{\bar \psi}_\mu \, P_R \, {\tilde \chi}^c_j \;\;\;,\;\;\;
\left( {\bar \psi}_\mu \, \left[ \gamma^\rho ,\, \gamma^\sigma \right] \gamma^\mu \, P_L \, {\tilde \chi}_j^c \right)^\dagger = - {\bar \psi}_\mu \, \left[ \gamma^\rho ,\, \gamma^\sigma \right] \gamma^\mu \, P_R \,{\tilde \chi}_j
\end{equation}

\section{Explicit derivation of the effective goldstino-matter Lagrangian in the non-derivative form}
\label{app:equivalence}

{\bf Statement.} {\it $
\left[\frac{\partial\mathcal{L}_\text{MSSM}}{\partial\Phi_i}-\partial_\mu
\frac{\partial\mathcal{L}_\text{MSSM}}{\partial(\partial_\mu\Phi_i)}
\right] \frac {\delta\Phi_i}\epsilon $
does not contribute to $S$-matrix elements, at all orders in perturbation theory (with the only restriction that no goldstino enters in propagators), for arbitrary initial and final state, with one goldstino external line.}\\

{\bf Proof.} Specifically, we need to show that
\beq
\langle f \vert {\rm T} \left\{ \exp \left[ i\int d^4 x
\mathcal{L}_\text{int.} \right]\int d^4y \left[ \frac{\partial
\mathcal{L}_\text{MSSM}}{\partial\Phi_i}-
\partial_\mu\frac{\partial \mathcal{L}_\text{MSSM}}{\partial(\partial_\mu\Phi_i)} \right]\frac{\delta\Phi_i}{\epsilon}\chi \right\} \vert i \rangle = 0 ,
\label{tobeproven}
\eeq
where $\mathcal{L}_\text{MSSM}
=\mathcal{L}_\text{free}+\mathcal{L}_\text{int.}$, $\epsilon$ is the
global SUSY variation parameter, $\Phi_i$ denotes any of the MSSM
fields, and $\chi$ is the goldstino. This proof is necessary to go
from eq. (\ref{eq:non-der-beforeproof}) to eq. (\ref{eq:non-der}) in
the main text.  The term with $\mathcal{L}_\text{free}$ includes the
free action for the MSSM fields.

We have
\beqn \frac{\partial
\mathcal{L}_\text{MSSM}}{\partial\Phi_i}-
\partial_\mu\frac{\partial \mathcal{L}_\text{MSSM}}{\partial(\partial_\mu\Phi_i)}&=&
\left( \frac{\partial \mathcal{L}_\text{free}}{\partial\Phi_i}-
\partial_\mu\frac{\partial \mathcal{L}_\text{free}}{\partial(\partial_\mu\Phi_i)}\right)+\left(
\frac{\partial \mathcal{L}_\text{int.}}{\partial\Phi_i}-
\partial_\mu\frac{\partial \mathcal{L}_\text{int.}}{\partial(\partial_\mu\Phi_i)} \right)\nonumber\\[3mm]
&\equiv& \text{ free e.o.m. of }\Phi_i+\left( \frac{\partial
\mathcal{L}_\text{int.}}{\partial\Phi_i}-
\partial_\mu\frac{\partial \mathcal{L}_\text{int.}}{\partial(\partial_\mu\Phi_i)}\right).
\label{EL}
\eeqn
The operator produced by the variation of $\mathcal{L}_\text{free}$ is the (classical) equation of motion
for the free field $\Phi_i$. For this reason, we denoted it as ``free e.o.m. of $\Phi_i$''. We actually prove that
\begin{eqnarray}
&&\langle f \vert {\rm T} \left\{ \exp \left[ i\int d^4 x
\mathcal{L}_\text{int.} \right] \int d^4y \left[
\text{ free e.o.m. of }\Phi_i \right]
\frac{\delta\Phi_i}{\epsilon}\chi
\right\} \vert i\rangle  \nonumber\\
&&=- \langle f \vert {\rm T} \left\{ \exp \left[ i\int d^4 x
\mathcal{L}_\text{int.} \right] \int d^4y \left[ \frac{\partial
\mathcal{L}_\text{int.}}{\partial\Phi_i}-
\partial_\mu\frac{\partial \mathcal{L}_\text{int.}}{\partial(\partial_\mu\Phi_i)} \right] \frac{\delta\Phi_i}{\epsilon}\chi
\right\}  \vert i\rangle  ,
\label{tobeproven2}
\end{eqnarray}
from which eq. (\ref{tobeproven}) immediately follows. We work out the l.h.s. of this expression, and we show that it is equal to the r.h.s.. When we use Wick's theorem to eliminate the time order product, the operator ``free e.o.m. of $\Phi_i$'' either acts on the initial or final state, or it is contracted with the field $\Phi_i$ inside $\exp \left[ i\int d^4 x \mathcal{L}_\text{int.} \right]$, present in the exponent. In the former case, one obtains zero, since the fields in the initial and final states are free fields, whose wave functions obey the free equations of motion. The contraction gives instead a nonvanishing contribution. We note that ``free e.o.m. of $\Phi_i$'' is linear in $\Phi_i$, so only one contraction with a single term in $\exp \left[ i\int d^4 x \mathcal{L}_\text{int.} \right]$ takes place. We therefore have (normal ordering is understood)
\begin{eqnarray}
{\rm l.h.s. \; of \; } (\ref{tobeproven2}) &=&
\langle f \vert T \left\{  \exp \Big[ i \, \int d^4 x \,
\contraction[2ex]{}{\mathcal{L}_\text{int.}}{\int d^4y \text{ free e.o.m. of }}{\Phi_i}
\mathcal{L}_\text{int.} \Big] \, \int d^4y \left[ \text{ free e.o.m. of }\Phi_i \right]
\frac{\delta\Phi_i}{\epsilon}\chi \right\}  \vert i\rangle \nonumber\\
&=& \langle f \vert T \left\{
 \sum_{n=1}^\infty \frac{1}{n!} \, \Big[ i \, \int d^4 x \,
\contraction[2ex]{}{\mathcal{L}_\text{int.}}{\int d^4y \text{ free e.o.m. of }}{\Phi_i}
\mathcal{L}_\text{int.} \Big]^n \, \int d^4y \left[ \text{ free e.o.m. of }\Phi_i \right]
\frac{\delta\Phi_i}{\epsilon}\chi \right\}  \vert i\rangle \nonumber\\
&=& \langle f \vert T \left\{   \sum_{n=1}^\infty\frac{[i\int
d^4x \mathcal{L}_\text{int.}]^{n-1}}{n!} \, n \, i \, \int d^4 z \,
\contraction[2ex]{}{\mathcal{L}_\text{int.}}{\int d^4y \text{ free e.o.m. of }}{\Phi_i}
\mathcal{L}_\text{int.}  \, \int d^4y \left[ \text{ free e.o.m. of }\Phi_i \right]
\frac{\delta\Phi_i}{\epsilon}\chi \right\}  \vert i\rangle \nonumber\\
\end{eqnarray}
where in the last step we have used the fact that ``free e.o.m. of $\Phi_i$'' contracts with all the $n$ actions appearing in the $n-$th term in the expansion series of the exponent. Since the sum in the last expression is again the expansion series of the exponent, we have found that
\begin{equation}
{\rm l.h.s. \; of \; } (\ref{tobeproven2}) = \langle f \vert T \left\{  \exp \left[ i\int d^4 x
\mathcal{L}_\text{int.} \right] \,  i \, \int d^4 z \,
\contraction[2ex]{}{\mathcal{L}_\text{int.}}{\int d^4y \text{ free e.o.m. of }}{\Phi_i}
\mathcal{L}_\text{int.}  \, \int d^4y \left[ \text{ free e.o.m. of }\Phi_i \right]
\frac{\delta\Phi_i}{\epsilon}\chi \right\}  \vert i\rangle
\label{proof-sep1}
\end{equation}

To proceed, we need to recall the specific dependence of the
interaction Lagrangian on $\Phi_i$. For MSSM fields, we have
\begin{equation}
\mathcal{L}_\text{int.} = \sum_n A_n \, \Phi_i^n + B^\mu \, \partial_\mu \Phi_i
= \sum_n A_n \, \Phi_i^n - \partial_\mu B^\mu \, \Phi_i
\label{phi-in-lint}
\end{equation}
where $n$ is an integer, and where the coefficients $A$ and $B^\mu$ can depend on fields other than
$\Phi_i$. Since the operator  ``free e.o.m. of $\Phi_i$'' does not contain any other field rather than $\Phi_i$, these coefficients do not participate to the contraction. Notice that in the last step we have disregarded a boundary term that does not contribute to the interaction action. Under the time ordering, we therefore have
\begin{eqnarray}
&&\int d^4 z \contraction[2ex]{}{\mathcal{L}_\text{int.}}{\int d^4y \text{ free e.o.m. of }}{\Phi_i}
\mathcal{L}_\text{int.}  \left( z \right) \, \int d^4y \left[ \text{ free e.o.m. of }\Phi_i \left( y \right)\right] \nonumber\\
&=& \int d^4 z \, \left[ \sum_n A_n \left( z \right) \, n \, \Phi_i^{n-1} \left( z \right) - \partial_\mu B^\mu \left( z \right) \right] \contraction[2ex]{}{\Phi_i}{\int d^4y \text{ free e.o.m. of }}{\Phi_i}
\Phi_i  \left( z \right) \, \int d^4y \left[ \text{ free e.o.m. of }\Phi_i \left( y \right)\right]
\label{inproof}
\end{eqnarray}

We can now use the fact that $\contraction{}{\Phi_i(z)}{\text{ free
e.o.m. of }}{\Phi_i(y)} \left[ \Phi_i(z) \text{ free e.o.m. of
}\Phi_i(y) \right] = i \, \delta^{(4)} \left( y-z \right)$. This is
immediate from the fact that the contraction of $\Phi_i(z)$ with the
operator $\Phi_i(y)$ entering in the expression in square
parenthesis is the propagator, which is the ``inverse'' of the
operator that forms the equation of motion (for instance,
$\square+m^2$, if $\Phi_i$ is a scalar). We also use the fact that
\begin{equation}
\sum_n A_n \left( z \right) \, n \, \Phi_i^{n-1} \left( z \right) - \partial_\mu B^\mu \left( z \right) =
\left[ \frac{\partial
\mathcal{L}_\text{int.}}{\partial\Phi_i}-
\partial_\mu\frac{\partial \mathcal{L}_\text{int.}}{\partial(\partial_\mu\Phi_i)}
\right]_z
\end{equation}
as can be immediately seen from eq. (\ref{phi-in-lint}). Therefore, eq. (\ref{inproof}) can be continued to give
\begin{equation}
\int d^4 z \contraction[2ex]{}{\mathcal{L}_\text{int.}}{\int d^4y \text{ free e.o.m. of }}{\Phi_i}
\mathcal{L}_\text{int.}  \left( z \right) \, \int d^4y \left[ \text{ free e.o.m. of }\Phi_i \left( y \right)\right]
= \int d^4 y \, i \, \left[ \frac{\partial
\mathcal{L}_\text{int.}}{\partial\Phi_i}-
\partial_\mu\frac{\partial \mathcal{L}_\text{int.}}{\partial(\partial_\mu\Phi_i)}
\right]_y
\end{equation}
Inserting this into eq. (\ref{proof-sep1}), we finally have
\begin{eqnarray}
{\rm l.h.s. \; of \; } (\ref{tobeproven2}) &=& \langle f \vert T \left\{ \exp \left[ i\int d^4 x
\mathcal{L}_\text{int.} \right] \,  i \, \int d^4 y \, i \, \left[ \frac{\partial
\mathcal{L}_\text{int.}}{\partial\Phi_i}-
\partial_\mu\frac{\partial \mathcal{L}_\text{int.}}{\partial(\partial_\mu\Phi_i)}
\right] \frac{\delta\Phi_i}{\epsilon}\chi \right\} \vert i\rangle \nonumber\\
&=& {\rm r.h.s. \; of \; } (\ref{tobeproven2})
\end{eqnarray}

This completes the proof.

\section{Amplitudes for $W^+ \, W^- \, \rightarrow \, $ gravitino $+$ neutralino}
\label{app:amplitudes}

The amplitude for the scattering  $W^+ \left( k \right) + W^- \left(
k' \right) \rightarrow {\tilde G} \left( p \right) + {\tilde
\chi}_i^{0} \left( q \right)$ can be written in the form
\begin{equation}
i \, {\cal M}_i =  i \left[ {\cal M}_{i,1\gamma} + {\cal M}_{i,1Z} + {\cal M}_{i,2} + {\cal M}_{i,3} + {\cal M}_{i,4}
+ {\cal M}_{i,5 H_1} + {\cal M}_{i,5 H_2} \right]
\label{eq:sum-M}
\end{equation}
The index $i$ on the total and the partial amplitudes denotes the neutralino mass eigenstate produced in the reaction. The numerical index on the partial amplitudes  denotes the order of the corresponding diagram in Figure \ref{fig:wwdiagrams}. Notice that the first and last diagram in the Figure correspond to two different terms in (\ref{eq:sum-M}).

In writing the partial amplitudes, we treat Majorana spinors as
explained in \cite{Denner:1992vza}. The Feynman rules for the
vertices with the gravitino are immediately obtained from the terms
listed in Appendix \ref{app:vertices-all}, after eq.
(\ref{eq:all-formal}). The MSSM vertices entering in the diagrams
are instead obtained from \cite{MSSM} and \cite{GH1986}
\begin{eqnarray}
{\cal L}_{\rm MSSM} &&\supset - i  e  A^\nu \left[ W^{\mu -} \left(
2 \partial_\mu  W_\nu^+ - \partial_\nu W_\mu^+ - \eta_{\mu \nu}
\partial_\alpha  W^{\alpha +} \right) - W^{\mu +} \left( 2
\partial_\mu  W_\nu^- - \partial_\nu W_\mu^- - \eta_{\mu \nu}
\partial_\alpha \, W^{\alpha -} \right)
\right] \nonumber\\
&&\!\!\!\!- i  g  \cos \theta_w  Z^\nu  \left[ W^{\mu -} \left( 2
\partial_\mu  W_\nu^+ - \partial_\nu W_\mu^+ - \eta_{\mu \nu}
\partial_\alpha  W^{\alpha +} \right) - W^{\mu +} \left( 2
\partial_\mu  W_\nu^- - \partial_\nu W_\mu^- - \eta_{\mu \nu}
\partial_\alpha
W^{\alpha -} \right) \right] \nonumber\\
&&-g \, W_\mu^- \, {\bar{\tilde \chi}}_j^c  \left[ O_{ij}^L \, P_L +
O_{ij}^R \, P_R  \right] \, \gamma^\mu \, {\tilde \chi}_i^{0}
+g \, W_\mu^+ \, {\bar{\tilde \chi}}_j  \, \gamma^\mu \, \left[ O_{ij}^{L*} \, P_L +
O_{ij}^{R*} \, P_R  \right]  \, {\tilde \chi}_i^{0} \nonumber\\
&&+g \, M_W \, W_\mu^+ \, W^{\mu-} \, \left[
\cos \left( \beta - \alpha \right) \, H_1^0 + \sin \left( \beta - \alpha \right) \, H_2^0
\right]
\end{eqnarray}
where
\begin{equation}
O_{ij}^L \equiv  - \frac{1}{\sqrt{2}} \,  N_{i4} \, V_{j2}^* + N_{i2} \, V_{j1}^* \;\;\;,\;\;\;
O_{ij}^R \equiv   \frac{1}{\sqrt{2}} \,  N_{i3}^* \, U_{j2} + N_{i2}^* \, U_{j1}
\end{equation}
and we remind the reader that the matrices $N,U,V$ are defined in eqs. (\ref{eq:def-nuv}).

We find
\begin{eqnarray}
i \, {\cal M}_{i,1 \gamma} &=& {\bar \psi}_\mu \left( p \right) \, \frac{-i}{4 \, M_P}
\, \left[ \slashed{p} + \slashed{q} ,\, \gamma^\sigma \right] \gamma^\mu   \nonumber\\
&&\cdot \left[ \left( \cos \theta_w \, N_{i1}^* + \sin \theta_w \, N_{i2}^* \right) P_L +
\left( \cos \theta_w \, N_{i1} + \sin \theta_w \, N_{i2} \right) P_R \right] v_i \left( q \right)
\, \frac{- i \, g_{\sigma \sigma'}}{\left( p + q \right)^2} \nonumber\\
&&\cdot e \, \left[ - i \left( 2 k + k' \right)_\beta \, \delta_\alpha^{\sigma'} +
i  \left( k + 2 k' \right)_\alpha \, \delta_\beta^{\sigma'} + i \left( k - k' \right)^{\sigma'} \, \eta_{\alpha \beta} \right] \epsilon_+^\alpha \left( k \right) \, \epsilon_-^\beta \left( k' \right)
\end{eqnarray}
\begin{eqnarray}
i \, {\cal M}_{i,1 Z} &=& {\bar \psi}_\mu \left( p \right)
\frac{-i}{M_P}   \Big\{ \Big[ M_Z \left( \cos \beta  N_{i3}^* - \sin
\beta  N_{i4}^* \right) \eta^{\mu \sigma} + \frac{1}{4}
  \left[ \slashed{p} + \slashed{q} , \gamma^\sigma \right] \gamma^\mu  \left( \cos \theta_w  N_{i2}^* - \sin \theta_w  N_{i1}^* \right)  \Big] P_L \nonumber\\
&&\quad\quad\quad\quad\;\; + \Big[ M_Z \left( \cos \beta  N_{i3} -
\sin \beta  N_{i4} \right) \eta^{\mu \sigma} + \frac{1}{4}
  \left[ \slashed{p} + \slashed{q} , \gamma^\sigma \right] \gamma^\mu  \left( \cos \theta_w  N_{i2} - \sin \theta_w  N_{i1} \right) \Big] P_R \Big\} \nonumber\\
&&\cdot v_i \left( q \right)
\, \frac{i \left[ - g_{\sigma \sigma'} + \frac{\left( p + q \right)_\sigma \, \left( p + q \right)_{\sigma'}}{M_Z^2} \right]}{\left( p + q \right)^2 - M_Z^2} \nonumber\\
&&\cdot g \, \cos \theta_w \, \left[ - i \left( 2 k + k' \right)_\beta \, \delta_\alpha^{\sigma'} +
i  \left( k + 2 k' \right)_\alpha \, \delta_\beta^{\sigma'} + i \left( k - k' \right)^{\sigma'} \, \eta_{\alpha \beta} \right] \epsilon_+^\alpha \left( k \right) \, \epsilon_-^\beta \left( k' \right)
\end{eqnarray}
\begin{eqnarray}
i \, {\cal M}_{i,2} &=& {\bar \psi}_\mu \left( p \right) \, \frac{-i}{M_P} \, \, \Bigg[ \left(
\sqrt{2} \, M_W \, \cos \beta \, U_{j2}^* \, \delta^\mu_\alpha + \frac{1}{4} \left[ \slashed{k} ,\, \gamma_\alpha
\right] \gamma^\mu \, U_{j1}^* \right) P_L  \nonumber\\
&&\quad\quad\quad\quad\quad+ \left(  \sqrt{2} \, M_W \, \sin \beta \, V_{j2} \, \delta^\mu_\alpha
+ \frac{1}{4} \, \left[ \slashed{k} ,\, \gamma_\alpha \right] \gamma^\mu \, V_{j1}  \right) P_R \Bigg]
\nonumber\\
&&\quad\quad\cdot \frac{i \left( \slashed{p}-\slashed{k}+M_{\tilde \chi_j} \right)}{\left( p - k \right)^2 - M_{\tilde \chi_j}^2} \, \left( - i g \right)
\left[ O_{ij}^L \, P_L + O_{ij}^R \, P_R \right] \gamma_\beta \, v_i \left( q \right) \, \epsilon_+^\alpha \left( k \right) \, \epsilon_-^\beta \left( k' \right)
\end{eqnarray}
\begin{eqnarray}
i \, {\cal M}_{i,3} &=& {\bar \psi}_\mu \left( p \right) \, \frac{-i}{M_P} \, \, \Bigg[ \left(
\sqrt{2} \, M_W \, \cos \beta \, U_{j2} \, \delta^\mu_\beta + \frac{1}{4} \left[ \slashed{k'} ,\, \gamma_\beta
\right] \gamma^\mu \, U_{j1} \right) P_R  \nonumber\\
&&\quad\quad\quad\quad\quad+ \left(  \sqrt{2} \, M_W \, \sin \beta \, V_{j2}^* \, \delta^\mu_\beta
+ \frac{1}{4} \, \left[ \slashed{k'} ,\, \gamma_\beta \right] \gamma^\mu \, V_{j1}^*  \right) P_L \Bigg]
\nonumber\\
&&\quad\quad\cdot \frac{i \left( \slashed{p}-\slashed{k'}+M_{\tilde \chi_j} \right)}{\left( p - k' \right)^2 - M_{\tilde \chi_j}^2} \, \left(  i g \right)
\gamma^\alpha \left[ O_{ij}^{L*} \, P_L + O_{ij}^{R*} \, P_R \right] \, v_i \left( q \right) \, \epsilon_+^\alpha \left( k \right) \, \epsilon_-^\beta \left( k' \right)
\end{eqnarray}
\begin{eqnarray}
i \, {\cal M}_{i,4} &=& {\bar \psi}_\mu \left( p \right)  \, \frac{i \, g}{4 \, M_P} \, \left[ \gamma_\alpha ,\, \gamma_\beta \right] \gamma^\mu \left( N_{i2}^* \, P_L + N_{i2} \, P_R \right) \, v_i \left( q \right) \, \epsilon_+^\alpha \left( k \right) \, \epsilon_-^\beta \left( k' \right)
\end{eqnarray}
\begin{eqnarray}
i \, {\cal M}_{i,5H_1} &=& {\bar \psi}_\mu \left( p \right) \, \frac{1}{M_P} \, \, \left[ \left(
\cos \alpha \, N_{i3}^* + \sin \alpha \, N_{i4}^* \right) P_L - \left(
\cos \alpha \, N_{i3} + \sin \alpha \, N_{i4} \right) P_R \right] \left( - i \right) \left( k + k' \right)^\mu
\nonumber\\
&&\quad\quad\cdot   v_i \left( q \right) \,
\frac{i}{\left( k + k' \right)^2 - M_{H_1^0}^2} \, i \, g \, M_W \, g_{\alpha \beta} \, \cos \left( \beta - \alpha \right) \, \epsilon_+^\alpha \left( k \right) \, \epsilon_-^\beta \left( k' \right)
\end{eqnarray}
\begin{eqnarray}
i \, {\cal M}_{i,5H_2} &=& {\bar \psi}_\mu \left( p \right) \, \frac{1}{M_P} \, \, \left[ \left(
- \sin \alpha \, N_{i3}^* + \cos \alpha \, N_{i4}^* \right) P_L - \left(
- \sin \alpha \, N_{i3} + \cos \alpha \, N_{i4} \right) P_R \right] \left( - i \right) \left( k + k' \right)^\mu
\nonumber\\
&&\quad\quad\cdot   v_i \left( q \right) \,
\frac{i}{\left( k + k' \right)^2 - M_{H_2^0}^2} \, i \, g \, M_W \, g_{\alpha \beta} \, \sin \left( \beta - \alpha \right) \, \epsilon_+^\alpha \left( k \right) \, \epsilon_-^\beta \left( k' \right)
\end{eqnarray}
where $\psi_\mu ,\, v_i ,\, \epsilon_+^\alpha ,\, \epsilon_-^\beta$
denote the gravitino vector spinor, the neutralino  spinors, and the
$W^+$ and $W^-$ polarization vectors, respectively (the spin and
polarization indices are understood).

\end{document}